\documentclass[final,2p,times,twocolumn]{elsarticle}

\usepackage[left=1.5cm,right=1.5cm,top=2cm,bottom=1.5cm]{geometry}

\usepackage{amssymb}

\usepackage[T1]{fontenc}

 \usepackage{lineno}

\usepackage[version=4]{mhchem}
\usepackage{textcomp}
\usepackage{xspace}
\usepackage{xcolor}
\usepackage{physics}
\usepackage{booktabs}
 \usepackage{placeins}
\usepackage{pdflscape}
\usepackage{siunitx}
\usepackage{rotating}

\usepackage{hyperref}
\hypersetup{
	colorlinks=true,         
	breaklinks=true,         
	urlcolor= orange!50!black,         
	linkcolor= orange,        
    citecolor=orange,     
    filecolor=black,      
    pagebackref=true,
}

\newcommand \wn {cm$^{-1}$\xspace}
\newcommand \pgo {PGOPHER\xspace}

\newcommand \ccc {\ce{C3}\xspace}
\newcommand \esi {\textbf{supplementary material}\xspace}

\newcommand \ax {$\Tilde{\mathrm{A}}\ ^1\Pi_u- \Tilde{\mathrm{X}}\  ^1\Sigma_g^+$\xspace}
\newcommand \xx {$\Tilde{\mathrm{X}}\  ^1\Sigma_g^+ - \Tilde{\mathrm{X}}\  ^1\Sigma_g^+$\xspace}
\newcommand \as {$\Tilde{\mathrm{A}}\ ^1\Pi_u$\xspace}
\newcommand \xs {$\Tilde{\mathrm{X}}\  ^1\Sigma_g^+$\xspace}
\newcommand \at {$\Tilde{\mathrm{A}}$}
\newcommand \xt {$\Tilde{\mathrm{X}}$}

\newcommand{\rev}[1]{\textcolor{black}{ #1}}

\journal{Journal of Molecular Spectroscopy}

\begin{document}

\begin{frontmatter}



\title{The Bending of \ccc: Experimentally Probing the $l$-type Doubling and Resonance}

\author[inst3]{Marie-Aline Martin-Drumel\fnref{label1}}
\author[inst1]{Qiang Zhang\corref{cor2}}
\author[inst2]{Kirstin D. Doney}
\author[inst3,inst4]{Olivier Pirali}
\author[inst3]{Michel Vervloet}
\author[inst5]{Dennis Tokaryk}
\author[inst6]{Colin Western}
\author[inst2]{Harold Linnartz}
\author[inst1]{Yang Chen}
\author[inst1]{Dongfeng Zhao\corref{cor1}}

\cortext[cor1]{dzhao@ustc.edu.cn}
\cortext[cor2]{qiangzhang@ustc.edu.cn}
\fntext[label1]{marie-aline.martin@universite-paris-saclay.fr}

\affiliation[inst3]{
            organization={Universit\'{e} Paris-Saclay, CNRS, Institut des Sciences Mol\'{e}culaires d'Orsay},
            city={Orsay},
            postcode={F-91405}, 
            country={France}}
            
\affiliation[inst1]{
            organization={CAS Center for Excellence in Quantum Information and Quantum Physics, and Hefei National Laboratory for Physical Sciences at the Microscale, University of Science and Technology of China},
            addressline={Hefei}, 
            city={Anhui},
            postcode={230026}, 
            country={P. R. China}}

\affiliation[inst2]{
            organization={Laboratory for Astrophysics, Leiden Observatory, Leiden University},
            addressline={PO Box 9513}, 
            city={RA Leiden},
            postcode={NL2300}, 
            country={the Netherlands}}

\affiliation[inst4]{
            organization={SOLEIL Synchrotron, AILES beamline},
            addressline={l'Orme des Merisiers, Saint-Aubin}, 
            city={Gif-sur-Yvette},
            postcode={F-91190}, 
            country={France}}

\affiliation[inst5]{
            organization={Department of Physics, University of New Brunswick},
            city={Fredericton},
            state={NB},
            country={Canada}}

\affiliation[inst6]{
            organization={School of Chemistry, University of Bristol},
            city={Bristol},
            country={United Kingdom}}

\begin{abstract}
\ccc, a pure carbon chain molecule that has been identified in different astronomical environments, is considered a good probe of kinetic temperatures through observation of transitions involving its low-lying bending mode ($\nu_2$) in its ground electronic state. The present laboratory work aims to investigate this bending mode with multiple quanta of excitation by combining recordings of high resolution optical and infrared spectra of \ccc produced in discharge experiments. 
The optical spectra of rovibronic (\ax) transitions have been recorded by laser induced fluorescence spectroscopy using a single longitude mode optical parametric oscillator as narrow bandwidth laser source at the University of Science and Technology of China. 36 bands originating from \xt(0$v_2$0), $v_2 = 0-5$, are assigned. 
The mid-infrared spectrum of the rovibrational $\nu_3$ band has been recorded by Fourier-transform infrared spectroscopy using a globar source on the AILES beamline of the SOLEIL synchrotron facility. The spectrum reveals hot bands involving up to 5 quanta of excitation in $\nu_2$.  
From combining analyses of all the presently recorded spectra and literature data, accurate rotational parameters and absolute energy levels of \ccc, in particular for states involving the bending mode, are determined.  
A single \pgo file containing all available data involving the \xt\ and \at\ states (literature and present study) is used to fit all the data.
The spectroscopic information derived from this work enables new interstellar searches for \ccc, not only in the infrared and optical regions investigated here but also notably in the $\nu_2$ band region (around 63 \wn) where vibrational satellites can now be accurately predicted. This makes \ccc a universal diagnostic tool to study very different astronomical environments, from dark and dense to translucent clouds. 
\end{abstract}




\end{frontmatter}



\section{Introduction} \label{sec:intro}
The bare carbon chain molecule propadienediylidene, \ccc, has for long been the object of considerable interest to astronomers and spectroscopists alike.  It is one of the few species for which astronomical observations have preceded laboratory identification. Puzzling, unidentified lines around 4051 {\AA} were first detected in cometary optical emission in 1881 by \citet{Huggins1882} then observed toward many cometary bodies \citep[$e.g.$,][]{Swings1941, Swings1942, McKellar1948}. These lines were successfully reproduced in the lab by \citet{Raffety1916} and \citet{Herzberg1942} but their molecular carrier remained elusive until the work of \citet{Douglas1951} who conclusively identified \ccc  through an isotope-substitution experiment. First actual rovibronic assignments of transitions in this band to transitions of the \ax system were delayed until the work of \citet{Gausset1963, Gausset1965}. Because of its lack of a permanent dipole moment, \ccc does not possess a pure rotational spectrum and is thus not accessible by radio astronomy. Instead, its electronic and vibrational spectra in the optical and infrared regions, respectively, provide useful spectroscopic approaches for detecting and tracing this simple molecule in the interstellar medium (ISM). Optical detections of the \ax band of \ccc have established the species as an ubiquitous component of diffuse interstellar matter \citep[$e.g.$, ][]{Maier2001, Roueff2002, Oka2003, Adamkovics2003, Welty2012, Schmidt2014} making it an ideal probe of the physical and chemical conditions in such environments.
In the work of \citet{Schmidt2014}, the most extensive rovibronic \ccc series detected in space to date, seven \at--\xt\ rovibronic bands in addition to the origin band were observed in absorption towards HD 169454, a reddened star (because of molecular extinction along a line of sight crossing one or more molecular clouds),  in a high-quality absorption spectrum.  Rovibronic series have also been reported in cometary spectra \citep[$e.g.$,][]{Cochran2002}. 
In the past decades, \ccc has also been detected in circumstellar shells and in the dense ISM  via detection of transitions in its asymmetric stretching ($\nu_3 \sim 2040$\,\wn) \citep[$e.g.$,][]{Hinkle1988} and bending ($\nu_2 \sim 63$\,\wn) \citep[$e.g.$,][]{Cernicharo2000,Giesen2001,Mookerjea2010} vibrational bands. 
It is particularly interesting to note that the observation of several rovibrational transitions involving the low-lying $\nu_2$ fundamental mode have been reported in the carbon-rich star IRC+10216 \citep{Cernicharo2000}, in the molecular cloud Sagittarius B2 \citep{Cernicharo2000,Giesen2001}, and in the star forming cores W31C and W49N \citep{Mookerjea2010}.  The unusually low frequency of the bending mode makes vibrationally excited \ccc a powerful probe of the kinetic temperature in various interstellar environments \citep{Cernicharo2000}. 

Besides the case of the 4051 {\AA} comet band system, astronomical observations have also prompted other laboratory studies on \ccc. Following the early work by \citet{Douglas1951} and \citet{Gausset1965}, dedicated experiments have been conducted, aiming for a comprehensive characterization of the \ax electronic band system of \ccc  at low- \citep[see, $e.g.$,][]{Weltner1966, Smith1987, Smith1988, Northrup1989, Northrup1990, Rohlfing1990, Balfour1994} and high- \citep[$e.g.$,][]{Merer1967, Izuha1995, Izuha1998, Tokaryk1997, Baker1997, Tanabashi2005, Zhang2005,Mookerjea2010, Chen2010, Chen2011, Schmidt2014,Haddad2014} resolution. The work of \citet{McCall2003} led to a reassignment of the R(0) transition of the \ax (000)--(000) band, in agreement with astronomical data \citep{Adamkovics2003}. Optical measurements have enabled the accurate description of the vibrational pattern in the electronic ground state (GS), notably the determination of the infrared inactive $\nu_1$ symmetric stretch band center ($\sim$ 1224 \wn) \citep{Zhang2005}.

In almost simultaneous studies, the $\nu_3$ fundamental vibrational band was observed in the laboratory \citep{Matsumura1988} and in space \citep{Hinkle1988}. The $\nu_2$ bending mode was investigated by far-infrared spectroscopy \citep{Schmuttenmaer1990} subsequently enabling its interstellar detection \citep{VanOrden1995,Cernicharo2000} which then prompted further high resolution laboratory investigations of the band \citep{Gendriesch2003, Breier2016}. The $\nu_1 + \nu_3$ combination band was also granted some interest: first measured in the laboratory by \citet{Krieg2013}, it was subsequently re-investigated by \citet{Schroeder2018} who extended the known stretching vibrational manifold, with levels involving up to seven quanta of excitation in $\nu_1$ and three quanta in $\nu_3$, thus probing the potential energy surface (PES) of this ``floppy'' molecule to high energies along the stretching coordinate and providing precise sets of molecular constants. In contrast, the $\nu_2$ bending vibrational manifold remains unstudied at this level of thoroughness with only a limited number of dedicated investigations. Relatively low-resolution stimulated-emission pumping (SEP) studies \citep[$e.g.$, ][]{Northrup1989,Northrup1990,Rohlfing1989:SEP}  have resulted in the observation of levels involving $v_2 = 0 - 34$ in the electronic GS while the rovibronic investigations of the \ax band system by \citet{Gausset1965} revealed bands involving up to $v_2 =4$. More extensive high-resolution data from direct absorption infrared studies, by probing hot bands in either $\nu_3$ \citep{Kawaguchi1989} or $\nu_1+\nu_3$ \citep{Krieg2013}, have enabled the detection of levels with up to (only) two quanta of excitation in $\nu_2$.  High resolution spectroscopic information for transitions involving higher quanta of excitation in $\nu_2$ is highly desirable to characterize the PES along the bending coordinate. Such data will also serve the astronomical community as the observation of vibrationally excited \ccc will allow to probe kinetic temperature of various interstellar environments. 

Even though an exhaustive review of the extensive literature on \ccc is beyond the scope of this paper, it is worth mentioning that because of its relevance not only in astrophysics but also in molecular physics, many other studies have been dedicated to this species. Many theoretical investigations were conducted on its potential energy surface \citep[e.g.][]{Jensen1992, Mladenovic1994, Schroeder2016} aiming specifically at establishing its equilibrium structure (and resolving the long-standing ambiguity linear \textit{vs.} bent; the former being the census to date) and providing reliable energy term values and vibrational constants for excited vibrational levels in the electronic GS. Experimentally, other electronic band systems have been investigated \citep[$e.g.$,][]{Saha2006,Sasada1991, Tokaryk1995,Lemire1989} while \ccc isotopologues have been the subject of optical \citep[][]{Clusius1954,Zhang2021} and infrared \citep{Moazzen1993} studies. The vacuum ultraviolet photoionization spectrum of \ccc was also reported \citep{Nicolas2006}. \citet{Weltner1989} and \citet{vanOrden1998} have published detailed reviews on the experimental works on \ccc as of the late 1990's. \ccc is also included in many astronomical models \citep[$e.g.$,][]{Hassel2011, Rousselot2001} and dedicated experiments have been conducted to investigate its reactivity with neutral species in space \citep[see][]{Kaiser2002}.

In this paper, we report a joint optical (\ax band) and infrared ($\nu_3$ band) investigation of \ccc in which we detected and assigned bands involving levels with up to 5 quanta of excitation in $\nu_2$.  Many of these bands are reported here for the first time. The combined analysis of these two data sets enables the accurate determination of the spectroscopic constants of the species as well as absolute rovibrational level energies for the $v_2=0-5$ levels in the electronic GS. From this, accurate far-infrared transitions are derived for ro-vibrational transitions involving the $\nu_2$ mode and its hot band sequences guiding reliable interstellar searches. The paper is organized as follows: in Section~\ref{sec:expt}, we describe the two experimental approaches used in this work; in Section \ref{sec:res}, the experimental results and spectral analysis are presented; in Section \ref{sec:prednu2}, a prediction of the line position of the $\nu_2$ rovibrational hot bands in the far-infrared is reported;  and in Section~\ref{sec:astro}, the astronomical implications of the present data are discussed.
The recorded line positions, experimental spectra, and fit files are available in the extensive \esi.

\section{Experimental methods} \label{sec:expt}

\subsection{Optical Spectroscopy}

The optical spectroscopic study of the \ax transition of \ccc has been performed  between 380 and 410 nm at the University of Science and Technology of China by using a laser induced fluorescence (LIF) setup that has been described in \citet{Zhang2017,Zhang2018,Zhang2020}. 
In this experiment, \ccc molecules are produced in a pulsed DC discharge nozzle using two flat-top stainless steel electrodes \citep{Motylewski1999}. 
A gas mixture of 0.3\,\% acetylene (C$_2$H$_2$) diluted in argon is introduced into the nozzle by a General Valve (Series 9, 0.5 mm orifice). 
High voltage pulses ($\simeq -2000$\,V, 20\,\textmu s, 10\,Hz) are applied to one of the electrodes while the other is grounded in order to produce an intense pulsed plasma.
The plasma containing \ccc molecules is then expanded and adiabatically cooled by collisions with the buffer gas. About 10\,mm downstream, the molecular beam is crossed perpendicularly by a laser beam, which suppresses the Doppler broadening in the recorded spectra. Fluorescent emission from laser excited \ccc molecules is collected by a lens system perpendicular to the laser beam, guided into a grating spectrometer (Zolix, 0.5\,m) and then detected by a photomultiplier tube (Hamamatsu, R928). 

A home-built single-longitude-mode optical parametric oscillator (SLM-OPO) is employed as the laser source \citep{Zhang2017:note}. In the present study, the signal output of the OPO is frequency-doubled in a KDP (KH$_ 2 $PO$ _4 $) crystal to obtain tunable radiation between 350 and 450\,nm. A small portion ($\sim$ 5\,\%) of the OPO signal output is injected into a wavelength meter (High Finesse, WS7-60) for calibration. The wavelength meter is calibrated with a stabilized He-Ne laser, providing a frequency accuracy of $\sim 0.002$\,\wn for the laser source.

Special care had to be taken in the measurement of the relatively weak hot vibronic bands. This is because in the supersonic jet the population of an excited vibrational level is much lower than that in the $v = 0$ level and, in most cases, the hot bands are overlapped with stronger fundamental vibronic bands. To overcome these difficulties, the spectrometer is used as a narrow band pass filter which helps to distinguish fluorescence of the hot bands from other overlapping bands. 
As the fluorescence from different upper states often results in different dispersed spectra, and since the bandpass of the monochromator is extremely narrow ($\sim$ 0.5 nm), it becomes possible to only detect the fluorescence emission from upper states involving specific GS hot bands. This was realized by selecting a dispersion wavelength of the monochromator corresponding to the fluorescence from the upper electronic state (itself optically-pumped from vibrationaly excited states of the \xt\ state) to the lowest allowed vibrational level of the GS.

A laser excitation spectrum is recorded by measuring the intensity of the fluorescence as a function of the continuously tuned laser wavelength. 
This technique yields a very good signal-to-noise (S/N) ratio and a spectral resolution of $\sim$\,0.02\,\wn (corresponding to a resolving power of $\sim$\,1,200,000) for the strong bands. Based on the line width of the rovibronic transitions, the accuracy of the extracted line positions from our spectra can confidently be assumed to be of about 0.002\,\wn for the bands arising from $v''_2 = 0, 1$ in the ground electronic state [with the exception of the \at(000)--\xt(000) band for which a 0.005\,\wn uncertainty is used] and 0.005\,\wn for the other hot bands (with $v''_2 \geq 2$). Absolute frequency accuracies may be subject to small shifts because of combining spectra from different wavelength regimes that are recorded in different experimental setups, but comparison with previous studies \citep[$e.g.$,][]{Gendriesch2003,Schmuttenmaer1990, Tokaryk1997} allows for reliable calibration (see later in section \ref{sec:methodo}).

\subsection{Infrared Spectroscopy}
 
The absorption spectrum of \ccc has been recorded in the $\nu_3$ asymmetric stretch region (around $2000$\,\wn /  5\,\textmu m) using an experimental set-up available on the AILES beamline of synchrotron SOLEIL previously used to investigate the far-infrared spectra of various reactive species \citep[see][]{Martin-Drumel2011, Martin-Drumel2012, Martin-Drumel2014}. The schematic representation of the discharge set-up together with its implementation on the AILES beamline is given in Figure 1 of \citet{Martin-Drumel2011}. We briefly describe in this paper the main characteristics of the set-up and the discharge conditions we used to optimize the signal of \ccc. The discharge cell consists of a 1.1\,m long Pyrex tube (13\,cm inner diameter) equipped with multipass optics (White-type) allowing for 24\,m of absorption pathlength. The cell is connected under vacuum to a Bruker Fourier-transform (FT) infrared spectrometer and is separated from it by two wedged \ce{CaF2} windows. A total of seven cell inlets are used: two are connected to the two water-cooled electrodes (separated by 70\,cm), four allow for gas injection (buffer gas or sample) at different locations in the cell, and one located at the center of the cell is connected to a vacuum system (250\,m$^{3}$/h Roots blower).
To probe the positive column of the plasma, one of the electrodes is connected to the high voltage while the second is grounded.
In the present work, \ccc is synthesized using a discharge of He (injected both through the electrodes and through the two gas inlets closest to the multipass optics) seeded with a small amount of \ce{CH4} (injected using the inlets located closer to the center of the cell).

Under our experimental conditions, we find that the abundance of \ccc is very sensitive to both the discharge current and the pressure of \ce{CH4} precursor. 
To optimize the production of \ccc, we  use a rapid scan optical fiber spectrometer (Ocean Optics) to monitor the 4051\,\AA\ emission band. The \ce{CH4} pressure is slowly increased until reaching the maximum of the visible emission while simultaneously adjusting the discharge current. In the optimum conditions, about 1\,A of current (at 1\,kV DC) drives the discharge through  a ballast resistor of 100\,$\Omega$. A total pressure of about 1\,mbar of a mixture of \ce{CH4} seeded in He is injected in the cell and a continuous flow of gas is maintained. 
To record the absorption spectra in the 5\,\textmu m region, we use the globar internal source of the Bruker FTIR installed on the AILES beamline of the SOLEIL synchrotron, a CaF$_2$ beamsplitter, and an Indium Antimonide (InSb) detector. These experimental conditions, together with the use of a bandpass optical filter, limit the bandwidth of our acquisition to the 1850--2100\,cm$^{-1}$ range. The unapodized spectral resolution is set to 0.004\,cm$^{-1}$ and 138 scans are co-added. Despite the effort to optimize the synthesis of \ccc in our cell and reduce the noise floor, the most intense lines of our spectra correspond to only about 5\,\% of signal absorption resulting in a SNR of 10 at best. 

Aside from \ccc, intense absorption lines of CO (possibly produced by reaction with residual H$_2$O) belonging to the $\Delta v = 1$ sequences (with $v'' =0-14$, the 4 -- 3 band being the strongest one) of the species in its electronic GS are observed over the full spectral range covered in this study (Figure S1 in the \esi). Even though these series of intense lines hinder identification of weaker series of \ccc lines, they enable spectral calibration by comparison with the accurate wavenumbers values from Refs. \cite{Wappelhorst1997,Schneider1990a, Schneider1990b}. 
After calibration, the frequency accuracy is estimated for each  \ccc line  based on its full-width at half-maximum and signal-to-noise ratio \citep{Davis2001:FT} and ranges from 0.0005\,\wn to 0.006\,\wn.
The rotational contour of the CO bands also allows for estimation of its rotational temperature (see Figure S2 in the \esi) from comparison with simulations using the \pgo software \citep{Western2017}. Under our experimental conditions, a rotational temperature of 500\,K is found in each CO band. Since the observation of CO hot bands is limited by the spectral coverage rather than the levels population, no vibrational temperature value can be asserted. 
Besides CO lines, a few weak transitions of the fundamental bending mode $\nu_2$ of H$_2$O 
are also detected \citep{Loos2017} and several lines are assigned to the R branch of the fundamental $\nu_3$ band of C$_2$H in the 1860--1895\,\wn range \citep{Kanamori1987} (Figure S1).

\subsection{Methodology \label{sec:methodo}}

As mentioned previously, the $\nu_2$ vibrational band lies in the far-infrared region and is significantly weaker than the $\nu_3$ band \citep{Jensen1992}; for these reasons, it remains challenging to measure the hot band sequences involving $v_2$ which would provide vibrational energies in the $v_2$ progression. 
Alternatively, such information can be derived from a combined rovibronic and rovibrational analysis of bands involving various quanta of $\nu_2$ excitation in the electronic GS. This approach is used in the present study exploiting \ccc spectra in the optical and mid-infrared region. 

Initial analyses of optical and infrared data have been performed independently, mostly exploiting combination differences.
Subsequently, all literature and newly measured rotationally-resolved transitions have been imported into the \pgo software \citep{Western2017} which has been used successfully  to analyze specific bands of \ccc  previously \citep[][]{Saha2006, Haddad2014, Schroeder2018, Zhang2021}.
Some features of this software have proven particularly powerful in the present study. First, all assignments can be treated as separate input files (for instance, one file per band) hence easing the treatment of large datasets. Then, the software allows for a relatively straightforward simultaneous treatment of rovibrational and rovibronic data. Last but not least, graphical representations of residuals, simulated bands, and term values plots is an invaluable tool to refine assignments and detect perturbations.
We thus perform a combined fit using \pgo of the presently recorded data together with all available rovibronic (\ax only) and rovibrational (all known bands in the \xt\ state) literature data allowing for the most complete spectroscopic description of the \ccc molecule to date. 

\subsection{Spectroscopy of \ccc} \label{sec:spectro}

As already mentioned in the introduction,  detailed descriptions of the spectroscopy of \ccc can be found in \citet{Gausset1965} and \citet{Rousselot2001}; here we recall some aspects pertinent to this work.
In the \xs electronic GS, as a result of the presence of a doubly-degenerate bending vibration, the species exhibits \textit{l}-type doubling. Levels with $v_2 > 0$ are split into multiple \textit{l}-states, with $l$ the vibrational angular momentum, such that \textit{l} = $v$, $v - 2$, ..., 0 or 1. Because the \ccc bending frequency is unusually small ($\omega_2 \sim 63$\,\wn), the $l$-type doubling is unusually large.
In addition, the \as electronic state is subject to Renner-Teller coupling, splitting the $v_2$ bending states into $K= \abs{l \pm \Lambda}$ components, where $K$ is the total vibronic angular momentum and $\Lambda$ the projection of the electronic orbital angular momentum onto the molecular axis of the molecule. Again, the very small bending frequency yields significant effects on \ccc spectroscopy, here resulting in a large Renner-Teller interaction.
A typical representation of splittings for the bending mode of linear molecules experiencing Renner-Teller interaction is given in Figure 9 of \citet{Gausset1965}. 
Because \ccc is linear, thus centrosymmetric, and contains identical nuclei of zero spin, all the antisymmetric levels have zero weight by spin statistics. In other words, half of the rotational levels are missing. For example, for $\Sigma^+_g$ states, only even-$J$ ($e$) levels exist while for a $\Sigma^+_u$ state, only odd-$J$ ($f$) levels exist. For other states, all $J$ levels exist but with alternate $e$/$f$ labels. 

In the following, each vibronic level is labeled $M$($v_1v_2v_3$) $N_s$ where $M$ is the electronic state (\xt\ or \at), $v_i$ ($i=1-3$) refers to the quanta of excitation in each vibrational level, $N$ is the Greek letter associated to the $K$ quantum number, and $s$ the symmetry of the state ($g$ or $u$). For instance, the \xt(020) $\Delta_g$ state corresponds to the $v_2=2$, $K=l=2$ vibrational level in the \xs electronic state; the \at(020) $\Phi_u$ state corresponds to the $v_2=2$, $K=3$ (hence $l=2$) vibrational level in the \as electronic state. When discussing rovibrational transitions in the \xs\ state, the \xt\ is often omitted.

\section{Results and discussion} \label{sec:res}
\begin{figure*}[ht!]
    \centering
    \includegraphics[width=\textwidth]{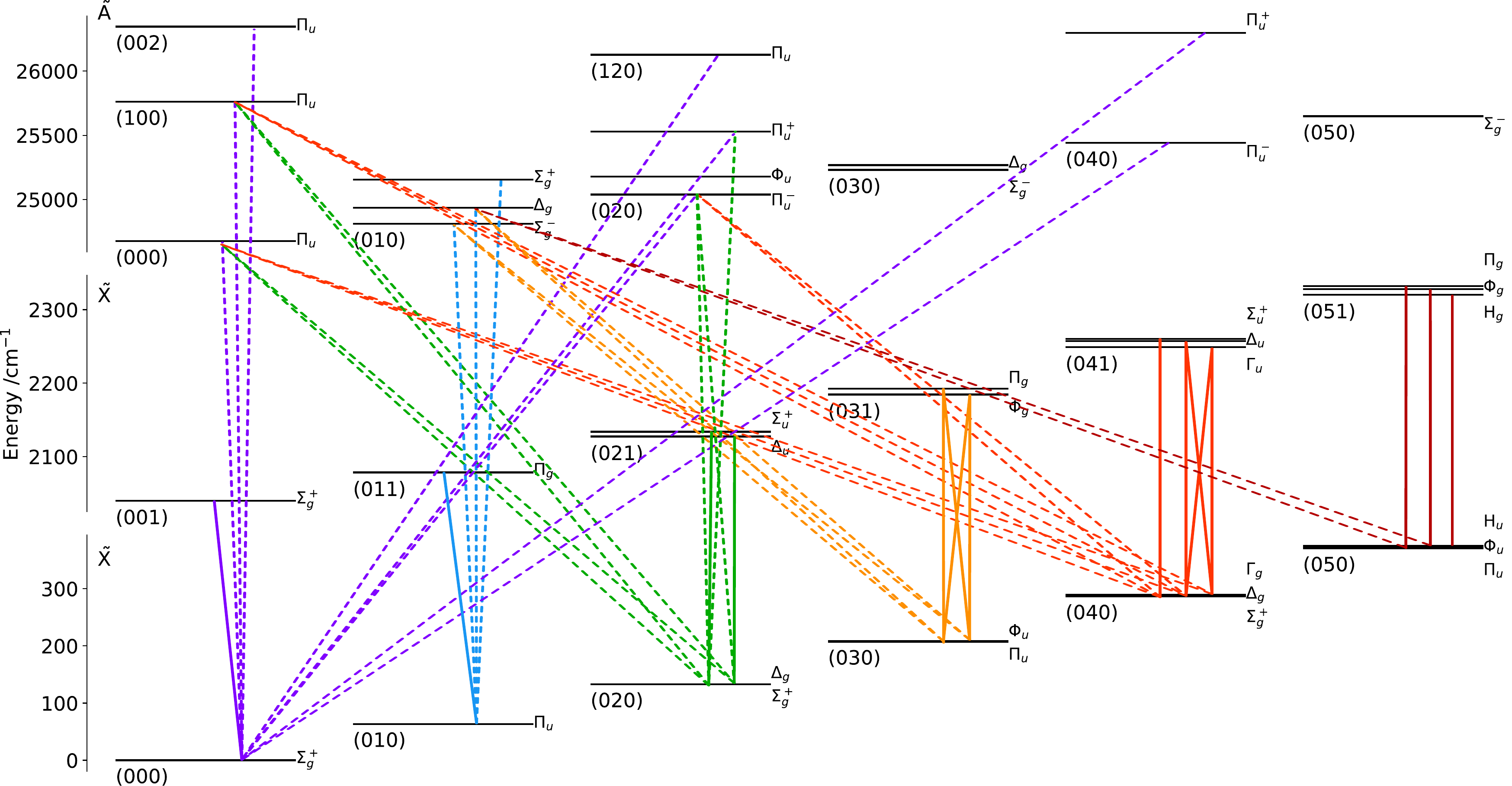}
    \caption{Schematic vibrational energy level diagram of \ccc together with optical (in dashed lines) and infrared (in plain lines) bands observed in this work. Energy levels are represented with increasing quanta of excitation in $v_2$ from left to right. Energies are from the combined fit performed in this work. Only the levels involved in bands observed in the present study are plotted, with the exception of the \at(0$v_2$0) levels, $v_2=0-5$, for which all levels included in the combined fit are shown. Transitions in the same color arise from the same \xt(0$v_2$0) lower state.   }
    \label{fig:C3bands}
\end{figure*}

Our rovibronic and rovibrational sets of data are extremely complementary as illustrated in Figure \ref{fig:C3bands}, allowing to retrieve both accurate energies and spectroscopic constants for levels involving $v_2 = 0-5$ in the \xs state. 
With regard to the \xt(0$v_2$0) vibrational levels with $v_2=0-5$, the single fundamental piece of information not accessible using our combined data set alone is the energy of the \xt(010) $\Pi_u$ state (see Figure \ref{fig:C3bands}); a value that has fortunately been determined accurately by \citet{Schmuttenmaer1990}.

\subsection{Optical data} \label{sec:optical RA}

\begin{figure*}[th!]
	\begin{center}
		\includegraphics[width=2\columnwidth]{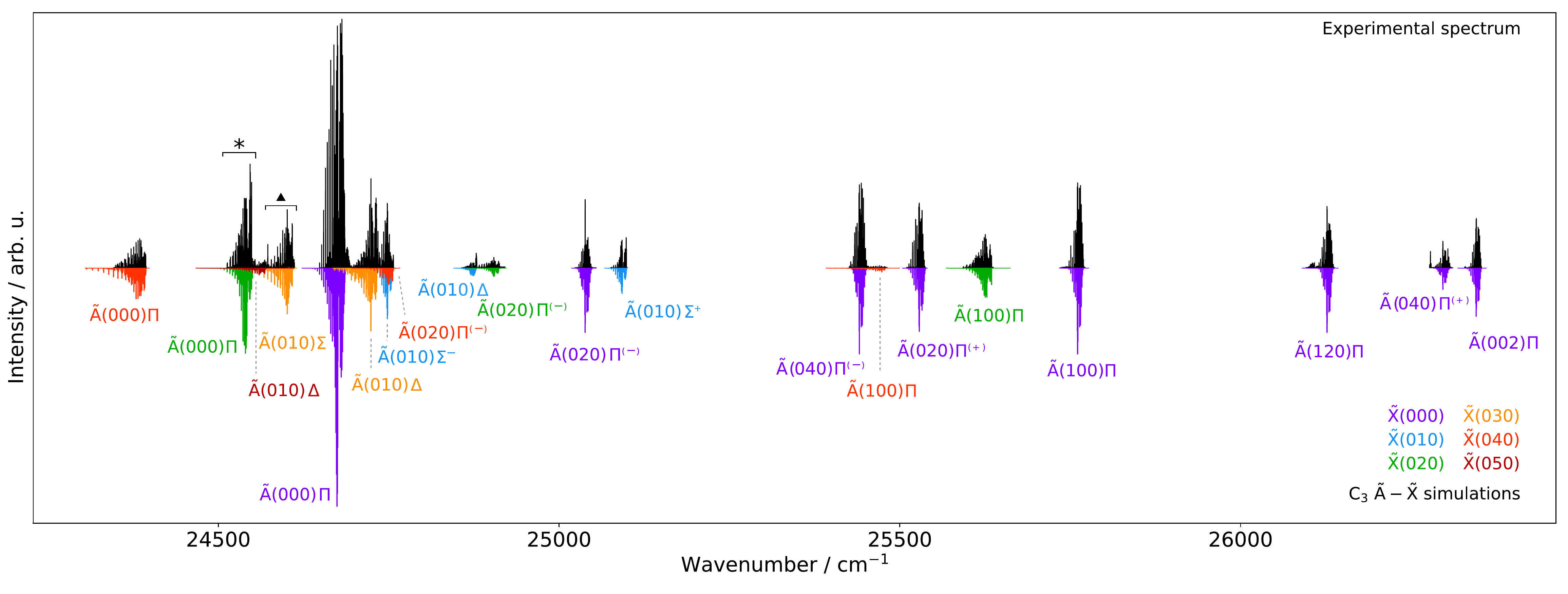}
		\caption{Overview of the electronic spectra of \ccc recorded in this work. \textit{Top traces:} Experimental spectra. \textit{Bottom traces:}  \pgo simulations (at thermal equilibrium and rotational temperatures reflecting each experimental band, from 20 to 150\,K; see Table S1 in the \esi). Intensities are in arbitrary units and the ratio between simulated bands is adjusted to reflect the experimental traces. Each simulated band is color coded according to the hot band of $\nu_2$ involved in the \xs state (the color sequence ranges from purple to red using the same color coding as in Figure \ref{fig:C3bands}). At this scale, bands involving different $l$ values in the \xt\ state, $e.g.$, \at(000)--\xt(020) $\Pi_u$--$\Sigma_g^+$ and $\Pi_u$--$\Delta_g$, are overlapping. Zooms onto the \at(000)--\xt(020) (highlighted by a star symbol on the figure) and \at(010)--\xt(030) $\Sigma_g^-$--$\Pi_u$ and $\Sigma_g^-$--$\Phi_u$ (highlighted by a triangle) band systems are presented in Figure \ref{fig:AX000-020} and \ref{fig:AX010-030}, respectively.
		}
		\label{Fig:C3_A-X_full}
	\end{center}
\end{figure*}

In the 24300--26400\,\wn region, 36 \ax vibronic bands have been recorded by LIF spectroscopy as summarized in Table~\ref{tab:OpticBandsDataset} and visible on Figure \ref{Fig:C3_A-X_full}. Of these, 17 are reported for the first time. 
The strongest bands in the optical spectra belong to transitions arising from the \xt(000) state (Figure \ref{Fig:C3_A-X_full}).
The \pgo software allows to estimate the rotational temperature of each band; it ranges for 20\,K to 150\,K (see Table S1 in the \esi). Typical linewidths reproducing the experimental spectra range from 0.02 \wn to 0.04 \wn. 
The present measurements are in good agreement with band values previously reported in the literature and often provide more accurate frequencies (see Table S2 in the \esi).  
Thanks to the relatively low rotational temperature combined with the high resolution resulting in well-resolved features, assignments of the new bands are relatively straightforward. Several bands appear heavily perturbed, in particular those involving the \at(020) $\Pi_u^{(-)}$ and \at(040) $\Pi_u^{(+)}$ states.


\setlength{\tabcolsep}{3pt}
\begin{table}[ht!]
    \centering \scriptsize
    \caption{\ax rovibronic  bands of the C$_3$ molecule observed in the present work.  $J''_\mathrm{max}$ values are reported for this work and the literature, together with the references of the previous works. When no literature value is reported, the band is observed for the first time in this work. Horizontal lines group bands arising from lower state levels with the same number of quanta of excitation in $\nu_2$. See Table S2 in the \esi for detailed information on the literature data and the present dataset. } 
    \label{tab:OpticBandsDataset}
\begin{tabular}{c r@{}c@{}l c c c c}

\toprule 
\multicolumn{4}{c}{Vibronic assignment} &  \multicolumn{1}{c}{Band Origin$^a$} & \multicolumn{1}{c}{This work} & \multicolumn{2}{c}{Literature} \\ \cmidrule{7-8}
\multicolumn{4}{c}{ } & \multicolumn{1}{c}{/ \wn} &\multicolumn{1}{c}{$J''_\mathrm{max}$} & \multicolumn{1}{c}{$J''_\mathrm{max}$} & \multicolumn{1}{c}{Refs.}\\
\midrule

\at(000)--\xt(000) 
    & $\Pi_u$&--&$\Sigma_g^+$ & 24676 & 24 & 64& \cite{Gausset1965,McCall2003,Zhang2005,Tanabashi2005,Haddad2014,Schmidt2014}\\
    & u$\Sigma_u$&--&$\Sigma_g^+\ ^b$ & 24679 & 14 & 16 &\cite{McCall2003,Zhang2005,Haddad2014}   \\
    & u$P_u$&--&$\Sigma_g^+\ ^b$    & 24676 & 6 & 8  &\cite{McCall2003,Zhang2005,Haddad2014}   \\
         
\at(020)--\xt(000) 
    & $\Pi_u^{(-)}$&--&$\Sigma_g^+$ & 25039 & 14 & 30 &\cite{Schmidt2014, Tokaryk1997} \\
    & $\Pi_u^{(+)}$&--&$\Sigma_g^+$ & 25529 & 18 & 50 &\cite{Gausset1965, Schmidt2014}\\

\at(040)--\xt(000) 
    & $\Pi_u^{(-)}$&--&$\Sigma_g^+$ & 25441 & 14 & 50& \cite{Gausset1965, Schmidt2014}\\
    & $\Pi_u^{(+)}$&--&$\Sigma_g^+$ & 26296 & 16 & 4 & \cite{Schmidt2014}\\
         
\at(002)--\xt(000) 
    & $\Pi_u$&--&$\Sigma_g^+$     & 26347 & 14 & 10 & \cite{Schmidt2014} \\ 
    
\at(100)--\xt(000) 
    & $\Pi_u$&--&$\Sigma_g^+$ & 25761 & 18 & 28& \cite{Gausset1965, Schmidt2014} \\
\at(120)--\xt(000) 
    & $\Pi_u$&--&$\Sigma_g^+$ & 26128 & 20 & \\

\midrule

\at(010)--\xt(010) 
    & $\Sigma_g^- $&--&$\Pi_u$ & 24749 & 17 & 47 &\cite{Gausset1965}\\
    & $\Delta_g$&--&$\Pi_u$    & 24872 & 18 & 51 &\cite{Gausset1965}\\   
    & $\Sigma_g^+ $&--&$\Pi_u$ & 25093 & 17 & 39 &\cite{Gausset1965}\\
\midrule

\at(000)--\xt(020) 
    & $\Pi_u$&--&$\Sigma_g^+$ & 24543 & 26 & 50 &\cite{Gausset1965}\\
    & $\Pi_u$&--&$\Delta_g$   & 24544 & 30 & 48 &\cite{Gausset1965}\\
    & u$\Sigma_u$&--&$\Sigma_g^+\ ^b$ & 25546 & 14 & \\
    & u$\Sigma_u$&--&$\Delta_g\ ^b$   & 25458 & 14 & \\
    & u$P_u$&--&$\Sigma_g^+\ ^b$      & 25542 & 6  & \\

\at(020)--\xt(020) 
    & $\Pi_u^{(-)}$&--&$\Sigma_g^+$ & 25906 & 26 & 28 &\cite{Tokaryk1997} \\
    & $\Pi_u^{(-)}$&--&$\Delta_g$ & 25906 & 25 & 30  & \cite{Tokaryk1997} \\

\at(100)--\xt(020) 
    & $\Pi_u$&--&$\Sigma_g^+$   & 25629 & 26 & \\
    & $\Pi_u$&--&$\Delta_g$     & 25629 & 23 & \\

\midrule

\at(010)--\xt(030) 
    & $\Sigma_g^- $&--&$\Pi_u$      & 24605 & 34 & 35& \cite{Gausset1965} \\
    & $\Sigma_g^- $&--&$\Phi_u$     & 24607 & 30 &  \\
    & $\Delta_g$&--&$\Pi_u$         & 24728 & 28 & 30 &\cite{Gausset1965} \\
    & $\Delta_g$&--&$\Phi_u$     & 24729 & 25 &  \\

\midrule

\at(000)--\xt(040) 
    & $\Pi_u$&--&$\Sigma_g^+$ & 24389 & 28 & 38 &\cite{Gausset1965} \\
    & $\Pi_u$&--&$\Delta_g$   & 24389 & 28 &   \\
    & $\Pi_u$&--&$\Gamma_g$   & 24391 & 30 &   \\
\at(020)--\xt(040) 
    & $\Pi_u^{(-)}$&--&$\Sigma_g^+$ & 24752 & 20 &\\
    & $\Pi_u^{(-)}$&--&$\Delta_g$   & 24752 & 21 &   \\
    & $\Pi_u^{(-)}$&--&$\Gamma_g$   & 24754 & 21 &   \\
\at(100)--\xt(040) 
    & $\Pi_u$&--&$\Sigma_g^+$   & 25475 & 24 &  \\
    & $\Pi_u$&--&$\Delta_g$     & 25475 & 24 &  \\

\midrule

\at(010)--\xt(050) 
    & $\Delta_g $&--&$\Pi_u$   & 24565 &  26 & \\
    & $\Delta_g $&--&$\Phi_u$  & 24656 &  22 & \\
\bottomrule

\end{tabular} 

\medskip
\begin{minipage}{0.9\columnwidth}
$^a$ Accurate values can be retrieved from the energies reported in Table S4.\\
$^b$ u$\Sigma_u$ and u$P_u$ are perturber states of the \at(000) state, unidentified so far (see text).   
\end{minipage}


\end{table}

\begin{figure}[th!]
    \centering
    \includegraphics[width=\columnwidth]{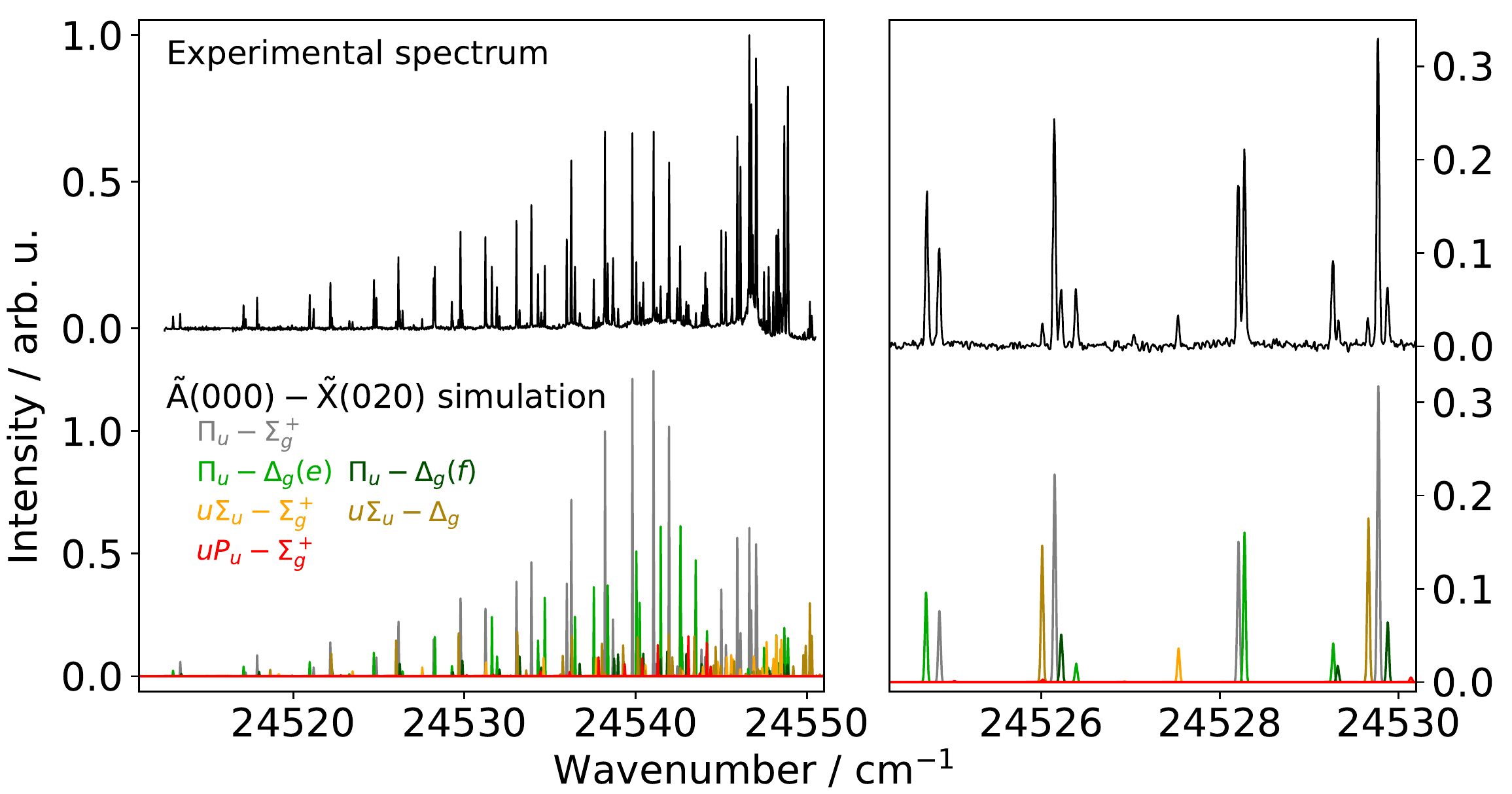}
    \caption{The \at(000)--\xt(020) band system observed around 24540 \wn (top traces) and comparison with a \pgo simulation at 70\,K (lower traces).
    (Left panel) Overview of the band; (right panel) zoom onto a portion of the spectral range. In the simulation, transitions within each sub-band are plotted in various colors. For the $\Pi_u$--$\Delta_g$ band, transitions involving even and odd $J''$ values ($e$ and $f$ levels in the lower state) are plotted in two shades of green with the intensity of the simulated $f$ symmetry divided by a factor 5 compared to the \pgo simulation. For the transitions involving the perturber states in the \as state, no scaling factor was used for the intensities as we could not define one that would be valid over the full spectral range.}
    \label{fig:AX000-020}
\end{figure}

In the \at(000)--\xt(020) band region, we assign for the first time transitions involving the u$\Sigma_u$ and u$P_u$ upper state perturbers, previously only observed in the \at(000)--\xt(000) band region \citep{McCall2003, Zhang2005, Haddad2014}. In total, 22 transitions are assigned in the bands involving \xt(020): 11 in the u$\Sigma_u$--$\Sigma_g ^+$ band (P-, Q-, R-branch; although the single Q-branch assignment remains tentative), 5 in the u$P_u$--$\Sigma_g^+$ band (P- and R-branch), and 6 transitions in the u$\Sigma_u$--$\Delta_g$ band (P- and R-branch). No transitions are observed for the u$P_u$--$\Delta_g$ branch. Figure \ref{fig:AX000-020} presents an overview of this band system with the sub-bands highlighted in various colors. These transitions are predicted by \pgo without setting up a transition moment for the corresponding bands (Table S1 in the \esi).
One can notice that the simulated intensities of the perturber transitions do not perfectly reproduce the experimental spectrum on Figure \ref{fig:AX000-020} (in particular, the u$\Sigma_u$--$\Delta_g$ band intensity is overestimated); we assume that some intensity perturbations are not properly taken into account. The agreement in line position, however, is quite satisfactory.
The present assignments thus confirm the assignment of the perturber lines  observed in the \at(000)--\xt(000) region.
It is also worth noticing that, on Figure \ref{fig:AX000-020}, the R-branches of the $\Pi_u$--$\Sigma_g ^+$ and $\Pi_u$--$\Delta_g$ bands (above 24545\,\wn) appear stronger in the experimental spectrum than on the simulation, and stronger than the Q-branches (the strongest features around 24540 \wn). Since no rotational temperature allows proper reproduction of such intensity ratios, this intensity difference is either the result of an experimental adjustment during the scan or of some discrepancy in the model that does not properly apply to this band system.
For the $\Pi_u$--$\Delta_g$ band, half of the $\Delta_g$ levels, with even $J''$ values, was previously observed by \citet{Gausset1965}. This peculiarity has led the authors to a better understanding of the $l$-uncoupling in the ground electronic state of \ccc. In the present study, we are able to assign transitions involving both even and odd $J''$ values. The transitions involving odd $J''$ values appear about 5 times weaker on the experimental spectrum than predicted using \pgo, and are thus significantly weaker than those involving even $J''$ values, which may explain why they remained unobserved thus far.

\begin{figure}[th!]
    \centering
    \includegraphics[width=\columnwidth]{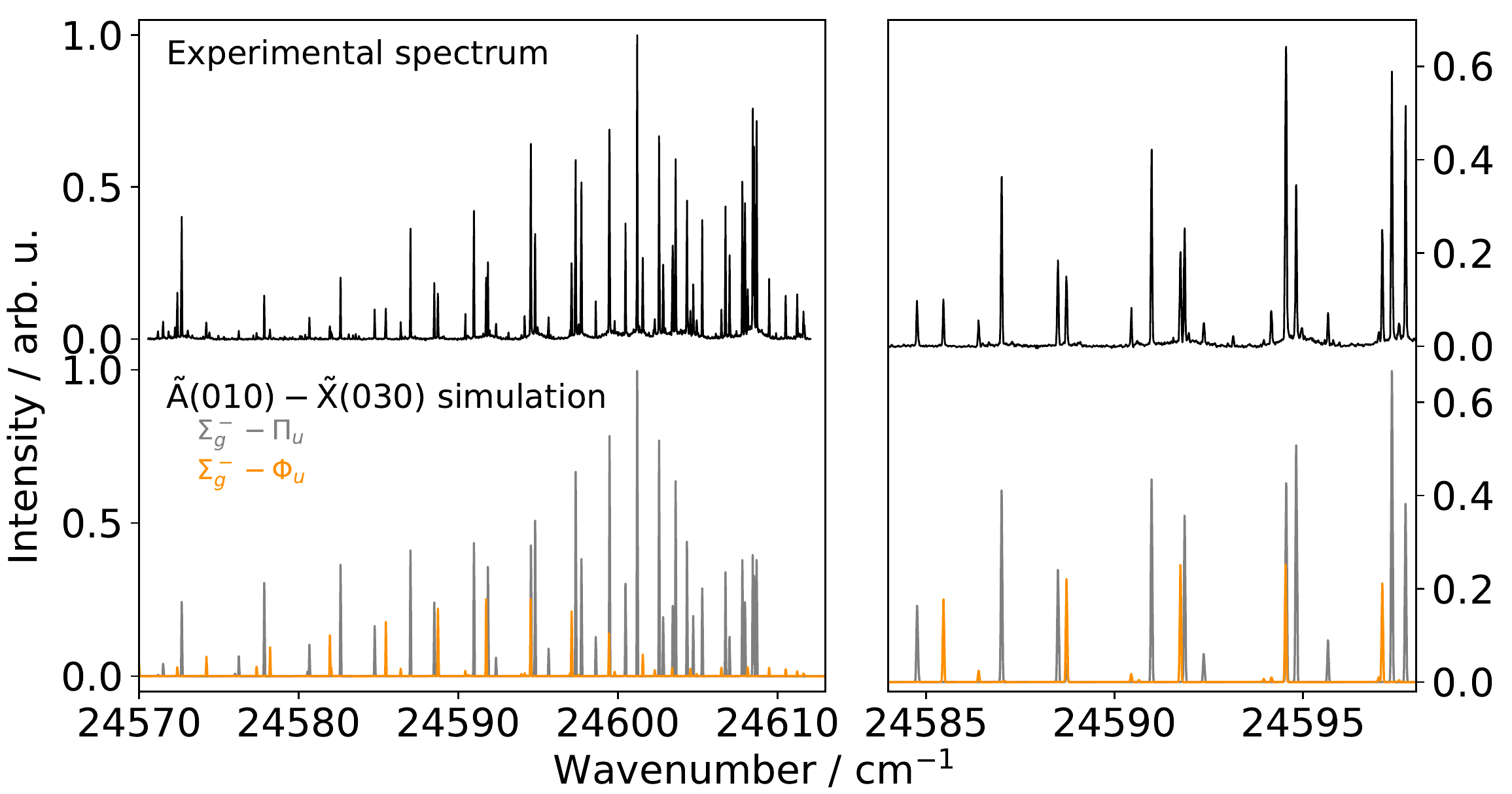}
    \caption{The  \at(010)--\xt(030) $\Sigma_g^-$--$\Pi_u$ and $\Sigma_g^-$--$\Phi_u$ band system observed around 24600 \wn (top traces) and comparison with a \pgo simulation at 150\,K (lower traces).
    (Left panel) Overview of the band; (right panel) zoom onto a portion of the spectral range. In the simulation, transitions from the $\Delta K= -1$ band are in gray and those from the $\Delta K= -3$ band are highlighted in orange.}
    \label{fig:AX010-030}
\end{figure}

Three $\Delta K= - 3$ bands are observed in this study, namely \at(010)--\xt(030) $\Sigma_g^-$--$\Phi_u$ (P-, Q-, and R-branches, Figure \ref{fig:AX010-030}), \at(000)--\xt(040) $\Pi_u$--$\Gamma_g$ (P- and Q-branches, tentative assignments in the R-branch, see Figure S3 in the \esi), and \at(020)--\xt(040) $\Pi_u^{(-)}$--$\Gamma_g$ (P- and Q-branches, Figure S4 in the \esi). To our knowledge, it is the first time that these transitions are observed for the \ax band.
These bands are spectrally intertwined with the corresponding $\Delta K= \pm 1$ bands arising from the same upper state and their intensity is properly predicted by \pgo  by perturbation ($i.e.$, even though no transition moment is used to predict them, see Table S1 in the \esi). These nominally forbidden transitions appear quite strong both on the simulation and the experimental trace, as visible on Figure \ref{fig:AX010-030}. We note, however, that again the R-branches of both bands (located above 24605\,\wn) are stronger on the experimental spectrum than predicted by the simulation.

Two bands of the \at(010)--\xt(050) band system, namely the $\Delta_g$--$\Pi_u$ and $\Delta_g$--$\Phi_u$ bands, are observed in this work (see Figure S5 in the \esi). Again, to the best of our knowledge, these bands are the first bands of the \ax transition probing the \xt(050) state at high resolution. A single band probing higher quanta of excitation in $\nu_2$ was previously reported in the literature, the \at(000)--\xt(060) $\Pi_u$--$\Sigma_g^+$ band for which Q-branches assignments were proposed by \citet{Merer1967}. In the present work, the spectral range where this band lies (around 24217\,\wn) has not been investigated and no other band involving the \xt(060) level is observed, preventing assignments confirmation by combination differences. 
Additional figures showing overviews of all \ax bands observed in this study are reported in the \esi (Figure S6--S17).

\subsection{Infrared data} \label{sec:IR RA}
\begin{figure*}[ht!]
	\begin{center}
		\includegraphics[width=2\columnwidth]{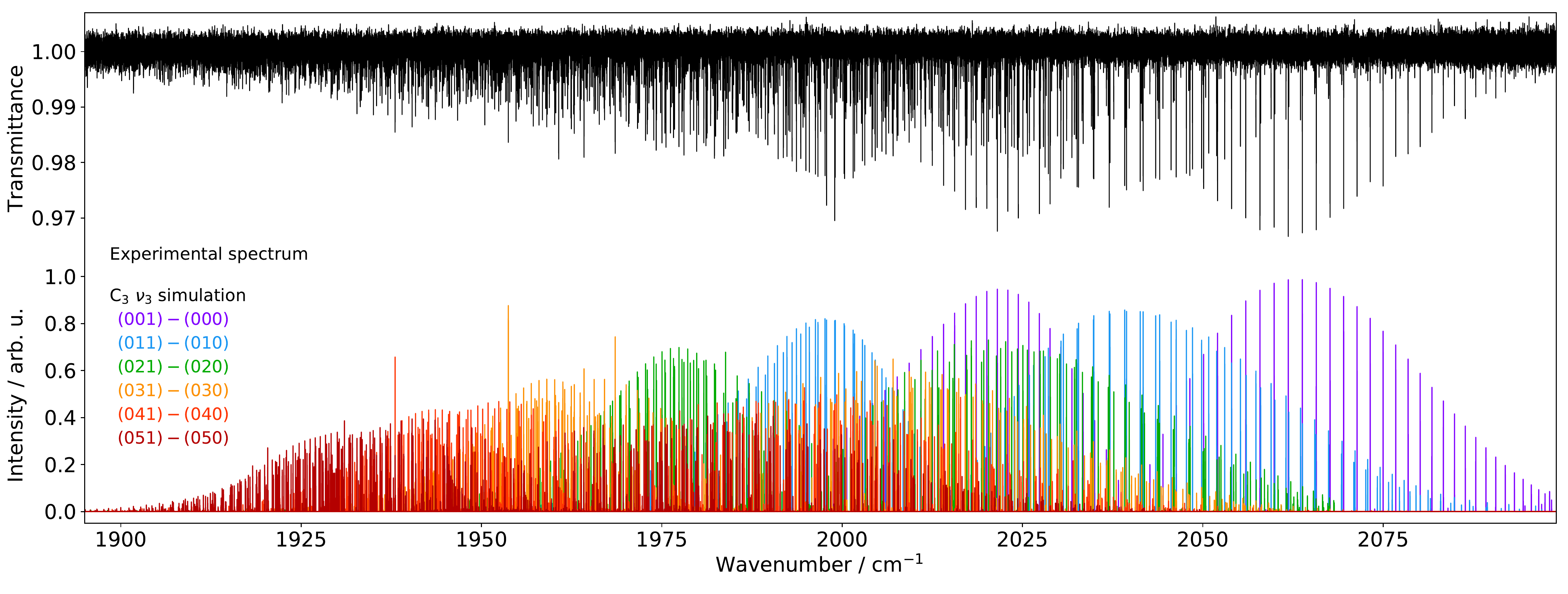}
		\caption{Spectrum of \ccc in the 1850--2100 cm$^{-1}$ spectral range. \textit{Top trace:} Absorption spectrum presented in transmittance. The intense absorption lines of CO as well as the lines of \ce{H2O} and \ce{C2H} have been removed from the experimental trace for clarity. \textit{Bottom trace:}  A 700 K \pgo simulation (at thermal equilibrium) of the $\nu_3$ band of \ccc and its hot bands involving up to five quanta of excitation in $\nu_2$. The hot band sequence is plotted in a color sequence ranging from purple to dark red (the color coding is the same as in Figure \ref{fig:C3bands}). The simulation is normalized to the strongest transition of the fundamental $\nu_3$ band.  
		}
		\label{Fig:C3_Full}
	\end{center}
\end{figure*}

\begin{table}[ht!]
	\scriptsize
	\centering
	\caption{Rovibrational bands of the C$_3$ molecule observed in the present work. $J''_\mathrm{max}$ values are reported for this work and the literature. When no literature value is reported, the band is observed for the first time in this work. Horizontal lines group bands arising from lower state levels with the same number of quanta of excitation in $\nu_2$. See Table S3 in the \esi for more information on the literature data and the present dataset.}
	\label{tab:MIRBandsDataset}

\begin{tabular}{c r@{}c@{}l c c  c c}

\toprule 
\multicolumn{4}{c}{Vibronic assignment} &  \multicolumn{1}{c}{Band Origin$^a$} & \multicolumn{1}{c}{This work} & \multicolumn{2}{c}{Literature} \\ \cmidrule{7-8}
\multicolumn{4}{c}{ } & \multicolumn{1}{c}{/ \wn} &\multicolumn{1}{c}{$J''_\mathrm{max}$} & \multicolumn{1}{c}{$J''_\mathrm{max}$} & \multicolumn{1}{c}{Refs.}\\
\midrule

\xt(001)--\xt(000) 
& $\Sigma_u^+$&--&$\Sigma_g^+$   & 2040 & 60 & 52 & \cite{Matsumura1988, Kawaguchi1989}\\

\midrule
\xt(011)--\xt(010) 
& $\Pi_g$&--&$\Pi_u$     & 2015 & 60 & 41 & \cite{Kawaguchi1989}\\

\midrule
\xt(021)--\xt(020) 
& $\Sigma_u^+$&--&$\Sigma_g^+$     & 2001 & 54 & 32& \cite{Kawaguchi1989}\\
& $\Delta_u$&--&$\Delta_g$     & 1994 & 55 & 32& \cite{Kawaguchi1989} \\
& $\Sigma_u^+$&--&$\Delta_g$    & 2003 & 20$^b$ & \\
& $\Delta_u$&--&$\Sigma_g^+$    & 1992 & 18$^b$ &  \\

\midrule
\xt(031)--\xt(030) 
& $\Pi_g$&--&$\Pi_u$     & 1985 &  51  & \\
& $\Phi_g$&--&$\Phi_u$   & 1976 &  55  & \\
& $\Pi_g$&--&$\Phi_u$    & 1988 &  23  & \\
& $\Phi_g$&--&$\Pi_u$    & 1973 &  31  & \\

\midrule
\xt(041)--\xt(040) 
& $\Sigma_u^+$&--&$\Sigma_g^+$   & 1973 & 48 & \\
& $\Delta_u$&--&$\Delta_g$       & 1970 & 49 & \\
& $\Gamma_u$&--&$\Gamma_g$       & 1960 & 49 & \\
& $\Gamma_u$&--&$\Delta_g$       & 1956 & 30 & \\
& $\Delta_u$&--&$\Gamma_g$       & 1974 & 28 & \\

\midrule
\xt(051)--\xt(050)
& $\Pi_g$&--&$\Pi_u$     & 1961 &  48  & \\
& $\Phi_g$&--&$\Phi_u$   & 1956 &  34  & \\
& H$_g$ &--& H$_u$       & 1945 &  44  & \\
\bottomrule
    \end{tabular}

\medskip
\begin{minipage}{0.9\columnwidth}
$^a$ Accurate values can be retrieved from the energies reported in Table \ref{tab:CtsX0v20}.\\
$^b$ Tentative assignment.
\end{minipage}
\end{table}

The full FTIR absorption spectrum of \ccc recorded in this work is shown in Figure \ref{Fig:C3_Full}. The intense absorption features from CO, H$_2$O, and C$_2$H have been removed from the experimental trace for clarity. 
The simulated spectrum given in the lower part of the figure is obtained using the \pgo software \citep{Western2017} using the band center and rotational constants obtained from our fit (see section \ref{sec:fit}), a Gaussian lineshape with a 0.005\,\wn width, and assuming a 700\,K rotational temperature (which was found to better reproduce the fundamental $\nu_2$ band; we note that this temperature is higher than that found for CO). The hot band sequence involving increasing values of $v''_2$ extends toward lower frequencies as visually indicated by the colored sequence. 
Figure \ref{Fig:C3_Full} also illustrates the strong overlap of all the bands observed in this work which causes most of the difficulties in the assignment process. 
For example, the lines arising from the (051)--(050) bands span over most of the spectral region. Since they are weak and hindered by many other more intense lines, their assignment was challenging. 
In total,  18 rovibrational bands of \ccc have been observed in this study, 14 of them for the first time (see Table \ref{tab:MIRBandsDataset}).
A detailed account on this work and available literature of rovibrational data is available in Table S3 in the \esi. In this work, the assignment of the infrared data was performed separately from the optical measurements presented in the previous sections and  strongly relied on the literature data (including optical studies).

\begin{figure}[ht!]
    \centering
    \includegraphics[width=\columnwidth]{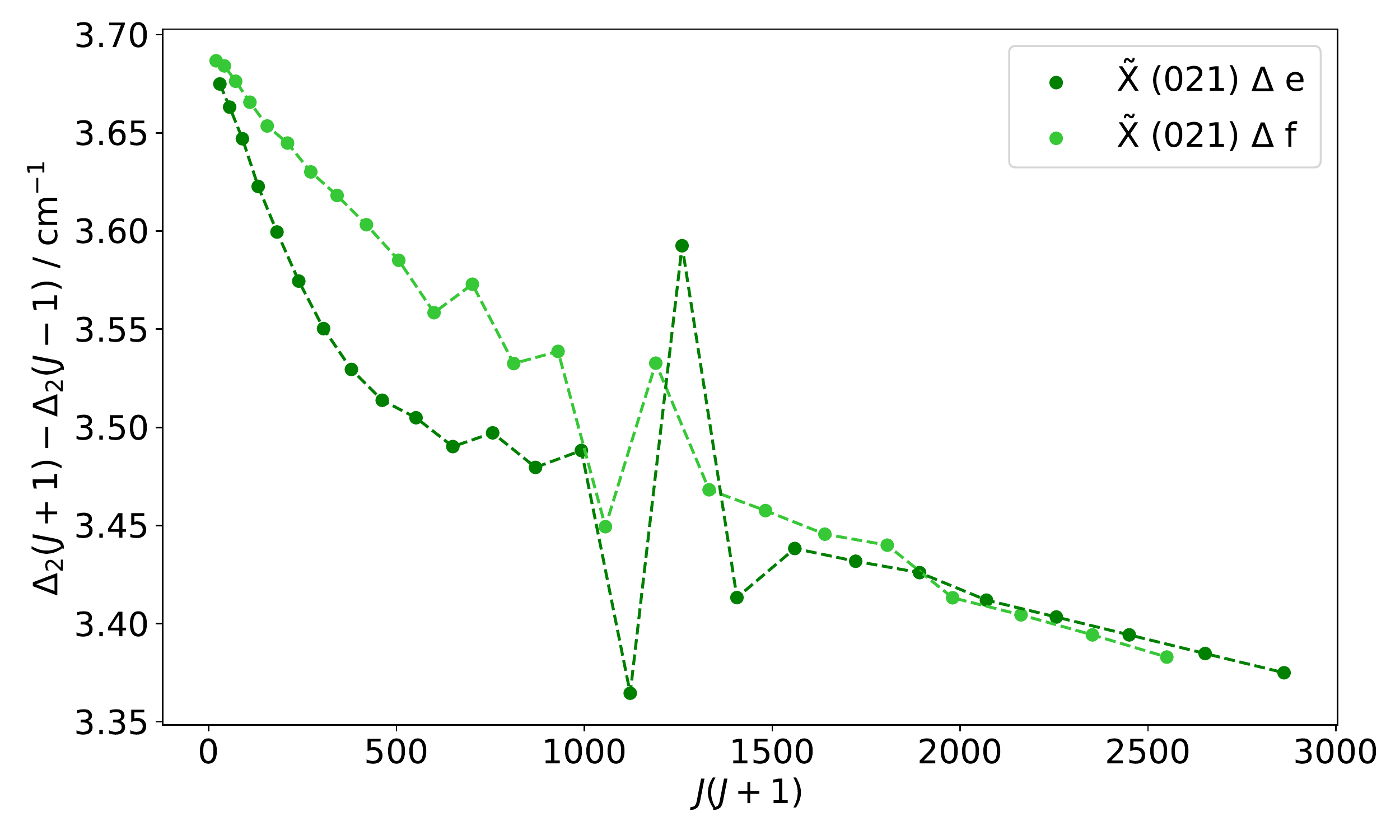}
    \caption{Perturbation analysis of the \xt(021) $\Delta_u$ $e$ and $f$  manifolds. }
    \label{fig:020FigPerturbs}
\end{figure}

The analysis of the (001)--(000) $\Sigma_u^+$--$\Sigma_g^+$, (011)--(010) $\Pi_g$--$\Pi_u$, and (021)--(020) $\Sigma_u^+$--$\Sigma_g^+$ and $\Delta_u$--$\Delta_g$ bands is rather straightforward  thanks  to  the  diode  laser  experiments  performed  by  \citet{Matsumura1988} and \citet{Kawaguchi1989}.  For all these bands, our observations are perfectly consistent with the literature measurements and allow for an extension of the dataset toward high-$J$ values (up to 55--60, see Table \ref{tab:MIRBandsDataset}). The frequency accuracy is slightly improved for the (001)--(000) and (011)--(010) bands (by up to a factor 2, with values as low as 0.0005 \wn) and similarly for the (021)--(020) bands (see Table S3 in the \esi).
Our extended dataset and the use of $\Delta_2^\mathrm{up}$($J$) (see the \esi for a detailed explanation) allowed to identify local perturbations in the upper \xt(021) $\Delta_u$ levels manifold. Figure \ref{fig:020FigPerturbs} shows the diagrams of the first derivative of the second differences, $\Delta_2^\mathrm{up}$($J$) values, calculated in the upper states. The highest shift occurs for the \xt(021) $\Delta_u$  $e$  manifold at $J'$ = 35 (the shift corresponds to about 10 times  the \ccc linewidth), and similar but smaller shifts are observed for the \xt(021) $f$  manifold. No notable level shifts are observed in (021) $\Sigma_g^+$ state. The  observed perturbations in the $\Delta_u$ state are likely caused by a Coriolis interaction with the \xt(190) $\Pi$ manifold, which mixes energy levels with $\Delta J$ = 0, $e \leftrightarrow e$ or $f \leftrightarrow f$, and $\Delta$\textit{l} = odd. While the \xt(190) $\Pi$ state has not yet been detected in SEP measurements, observations of \xt(180) $\Sigma$ at about 1993\,\wn puts the \xt(190) $\Pi$ state in the right energy range \citep{Northrup1990}.

We report for the first time numerous infrared transitions involving the (031)--(030), (041)--(040), and (051)--(050) hot bands. The assignment procedure relied mainly on combination differences using the work of \citet{Gausset1965} (see the \esi for a detailed explanation on the procedure). 
As no combination differences are available for the (030) $\Phi_u$ state, we used the results of successive fits that included $l$-type resonance with the $\Pi$ state to secure the assignments. Further confirmation is obtained by GS combination differences using the present optical data.
SEP data from \citet{Rohlfing1989:SEP} and \citet{Northrup1989}, despite their limited resolution, have provided crucial information on states for which little was known so far (vibrational energies and estimated $B$ values). In several cases, for example the (051)--(050) band manifold, this provided sufficient information to initiate an analysis.  

\citet{Rohlfing1989:SEP}  reported the vibrational energy of the \xt(051) $\Pi_g$ state [2330.9(5)\,cm$^{-1}$] as well as an estimation of the rotational constants for the lowest $J$ values [$B$ = 0.4718(13)\,cm$^{-1}$]. Using this information, we started our analysis for the $e$ and $f$  states. Once $\Pi_g$--$\Pi_u$ transitions were found (up to $J''$ = 40 for both $e$ and $f$ transitions), we used the $l$-type resonance to secure the assignments of $\Phi_g$--$\Phi_u$ and $\mathrm{H}_g$--$\mathrm{H}_u$ bands. 
The intensities of the P branches involving high $J$ values are very weak (red-end of our spectrum) and, as for the $\Delta_g$ states of (021), some Coriolis-type resonances seem to complicate the assignments. 
Figures S18--S23 in the \esi show the different infrared band manifolds observed in this study.

\begin{figure}[ht!]
    \centering
    \includegraphics[width=\columnwidth]{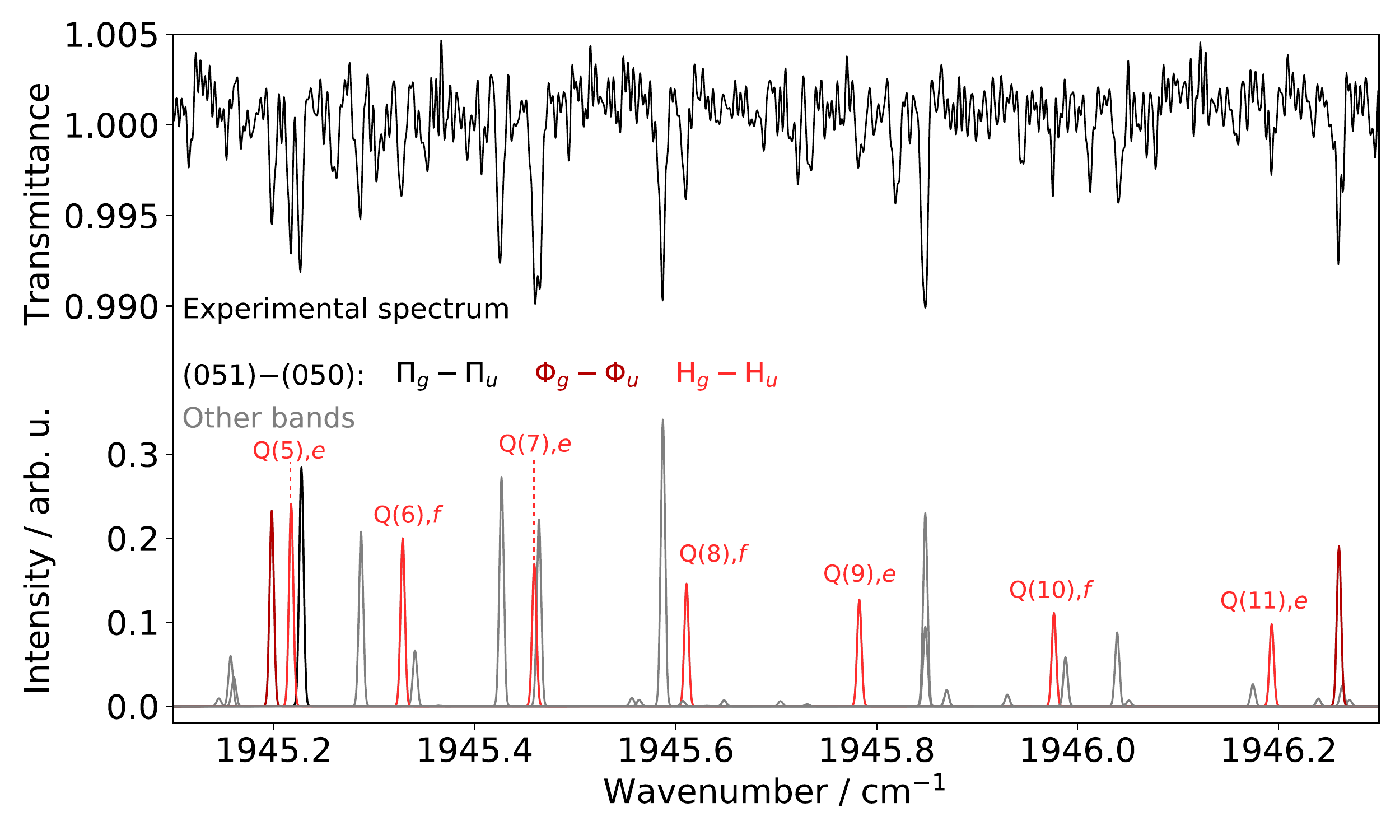}
    \caption[Q-branch in \xt(051)--\xt(050)  $\mathrm{H}_g$--$\mathrm{H}_u$ band]{Zoom onto the Q-branch of the \xt(051)--\xt(050)  $\mathrm{H}_g$--$\mathrm{H}_u$ band. \textit{Top trace:} experimental spectrum, in transmittance, after removal of CO, \ce{H2O}, and \ce{C2H} lines. \textit{Bottom trace:} \pgo simulations at 700 K (final set of parameters).}
    \label{fig:Q050}
\end{figure}

The rotational assignments proposed in this study are confirmed by the observation of several Q-branches, $i.e.$, no shift by one or more quanta of $J$ is possible in the proposed assignments. Such a $Q$-branch is shown on Figure \ref{fig:Q050} for the (051)--(050)  $\mathrm{H}_g$--$\mathrm{H}_u$ band, another example is provided for the (041)--(040)  $\Gamma_u$--$\Gamma_g$ band in Figure S24 in the \esi. Overall, Q-branch transitions were observed, for both $e$ and $f$ levels, for the (011)--(010) $\Pi_g$--$\Pi_u$ [Q(1),Q(3)], (021)--(020) $\Delta_u$--$\Delta_g$ [Q(2), tentative], (031)--(030) $\Phi_g$--$\Phi_u$ [Q(3)--Q(6)], (041)--(040) $\Gamma_u$--$\Gamma_g$ [Q(4)--Q(7)], and (051)--(050) $\Phi_g$--$\Phi_u$ [Q(3)--Q(5), Q(6) tentative] and $\mathrm{H}_g$--$\mathrm{H}_u$ [Q(5)--Q(11)] bands.

\begin{figure}[ht!]
    \centering
    \includegraphics[width=\columnwidth]{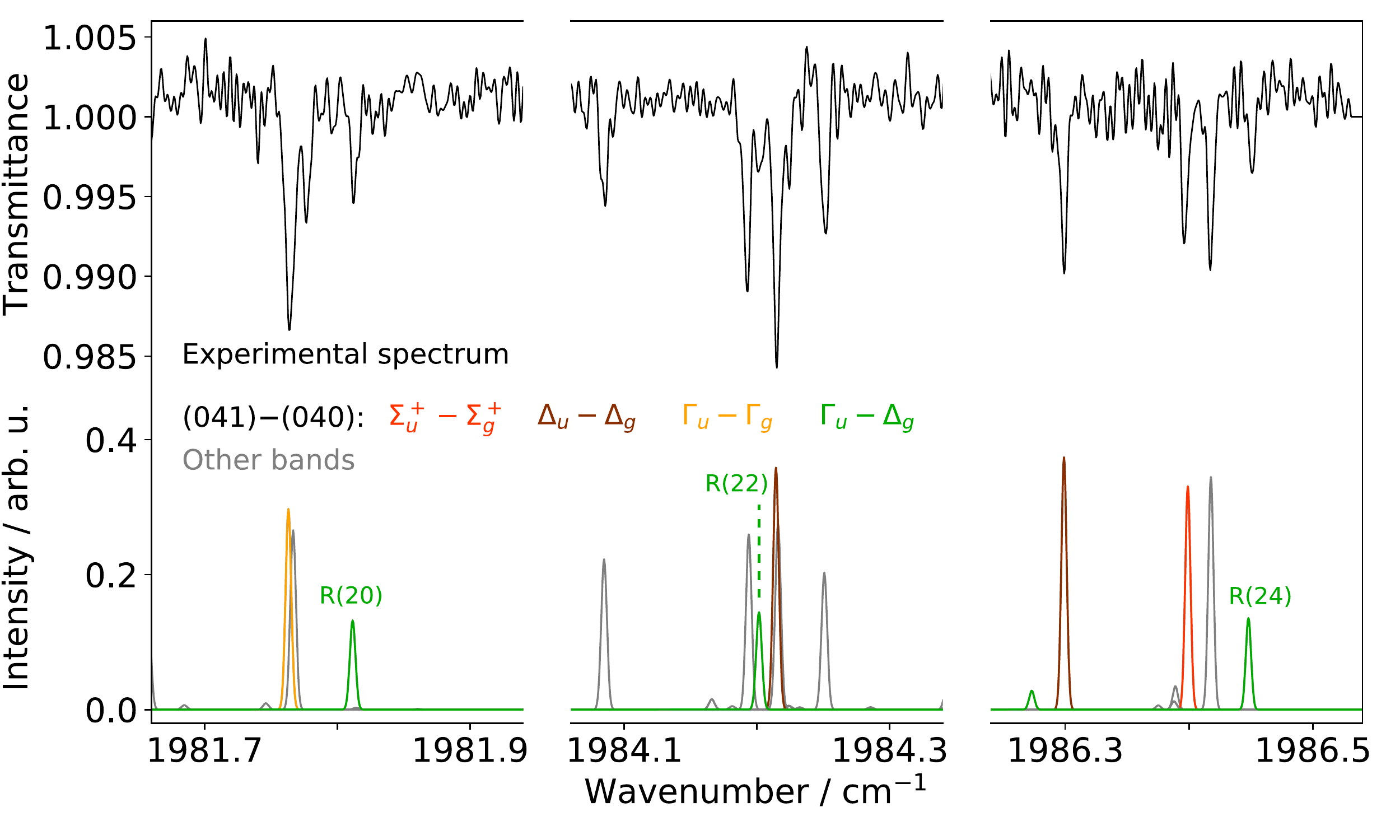}
    \caption{Zoom onto three consecutive R-branch transitions of the (041)--(040) $\Gamma_u$--$\Delta_g$ band. \textit{Top trace:} experimental spectrum, in transmittance, after removal of CO, \ce{H2O}, and \ce{C2H} lines. \textit{Bottom trace:} \pgo simulations at 700 K (final set of parameters,  normalized to the strongest transition of the fundamental $\nu_3$ band) where the $\Gamma_u$--$\Delta_g$ transitions are highlighted in green. The rest of the simulated transitions of the (041)--(040) bands visible in this range are plotted in shades of red while the other \ccc bands are simulated in gray.}
    \label{fig:040-forbidden}
\end{figure}

Once all the data were included in \pgo, we have also been able to assign some transitions involving $\Delta l = \pm 2$ bands. 
An interesting feature of \pgo is that it predicts these transitions without inputting a corresponding transition moment for these nominally ``forbidden'' transitions (similarly to what was described previously for the electronic spectra). 
For the (021)--(020) bands, the $\Delta l = \pm 2$ bands are predicted with relatively low intensities; and only features with relatively poor SNR are observed on the experimental spectrum. Only tentative assignments are made (7 for the $\Delta_u$--$\Sigma_g^+$ and 4 for the $\Sigma_u^+$--$\Delta_g$ band) and these are not included in the fit. 
For hot bands involving higher quanta of excitation in $\nu_2$, however, these predicted $\Delta l = \pm 2$ transitions have significant intensity, in particular for high $l$ values. Indeed, several features with reasonable SNR are observed on the spectrum as visible on Figure \ref{fig:040-forbidden} for the (041)--(040) $\Gamma_u$--$\Delta_g$ band. Additional examples are provided in the \esi for the (031)--(030) $\Pi_g$--$\Phi_u$ and $\Phi_g$--$\Pi_u$ bands (Figure S25) and the (041)--(040) $\Delta_u$--$\Gamma_g$ (Figure S26).
These ``cross-ladder'' transitions (if we refer to levels of a given $l$ value for a specific state as a ladder) provide constraints on the energy difference between the various $l$ levels of the vibrational levels for which they are observed.

In the high frequency part of the spectrum, most of the transitions can be assigned to the \ccc molecule (see for instance figure \ref{fig:040-forbidden} and figure S27 in the \esi). Toward the lower end of the spectrum, however, many transitions remain unassigned. These transitions probably arise from the R-branches of the (061)--(060) bands but assigning this  spectrum has not been possible despite our best efforts, mainly because the signal-to-noise ratio is rather poor in that region and spectroscopic assignments are challenging on the basis on a single branch.

\subsection{Combined fit} \label{sec:fit}

\subsubsection{Dataset}
The unique feature of the present work is that it becomes possible to combine the optical and infrared data sets. Available data from the literature and the present work for the \ax rovibronic and all \xx rovibrational transitions are listed in Tables S2 and S3 in the \esi. These detailed tables contain the $J_\mathrm{max}$ values, frequency uncertainties used in the combined fit, and the frequency offset eventually applied to the data. All available \ax data and rovibrational transitions in the electronic GS  ($i.e.$, data with list of assignments provided in the literature)  are included in the present fit with one exception, namely the work of \citet{Balfour1994} who reported the spectroscopic assignments for five \ax transitions. One of the rovibronic band observed by \citet{Balfour1994}, the \at(020)--\xt(000) $\Pi_u$--$\Sigma_g^+$ band, was re-investigated by \citet{Tokaryk1997} who proposed  a different spectroscopic assignment. In the present study, our assignments are in line with those of \citet{Tokaryk1997}. Additionally, we propose in this work an alternative spectroscopic assignment for the \at(002)--\xt(000) $\Pi_u$--$\Sigma_g^+$ band compared to \citet{Balfour1994}. Finally, when performing the combined fit, we noticed that the spectroscopic assignments proposed by \citet{Balfour1994} in the \at(200)--\xt(000) $\Pi_u$--$\Sigma_g^+$ band are incompatible with those of the \at(200)--\xt(200) $\Pi_u$--$\Sigma_g^+$ from \citet{Merer1967}. Since assignments from \citet{Merer1967} have proven consistent with literature data for the other bands they reported, we decided to only include the Merer values. 
There is no literature data able to confirm or infirm the spectroscopic assignments of the remaining two bands observed by \citet{Balfour1994}; at this stage we have chosen not to include these data in our fit.

As previously noted by \citet{Saha2006}, a serious difficulty arises in high resolution combination fits when rovibronic data are included from different light sources, as (small) spectral offsets are intrinsic to this approach.
This can be overcome by shifting the frequencies of one dataset by this offset value. It is often challenging, however, to determine which dataset presents an offset from the absolute transitions rest frequencies. The impact of this issue is relatively limited because only the absolute energy of the upper state is affected while the accuracy of the rotational constants and overall fit is not. In the present study, small offsets (typically of the order of several hundredths of a \wn)  are applied to several rovibronic datasets (see Table S2 for a detailed list of concerned data and offset values). We use the frequency offsets established by other authors from the literature when available (for instance, an offset of $+0.04$ \wn was determined for the data of \citet{Gausset1965} by \citet{Tanabashi2005}) for consistency reasons. When not available in the literature, we determine these offsets ourselves. 
Overall, offsets values range from 0.02 to 0.12 \wn, hence the absolute energies in the \as state may be affected by these amounts. 

As much as possible, data are included in the fit at their experimental accuracy (see Table S2 for typical values for each dataset). When no frequency accuracy was provided, we assume a value based on the dispersion of the frequencies from our best model. In some instances, the literature data are provided with an upper limit for the frequency error but the residuals from the fit show that this value is over-estimated. For example, the uncertainty on line frequency is assumed to be better than 0.01 \wn for the \at(000)--\xt(000) transitions reported by \citet{Zhang2005} while the residuals from our fit show that the line accuracy is probably more of the order of 0.005 \wn; that value is thus used in the present fit.
It has proven challenging to treat such a large dataset for one molecule subject to clear perturbations with transitions significantly deviating from the fit. 
Transitions severely perturbed are excluded from the fit while transitions slightly diverging are kept in the fit but with an increased frequency error (hence a lower weight), typically by a factor 10, in order to maintain a global $5\sigma$ deviation for the combined fit. Overall, the dataset contains 4425 observations (3957 rovibronic and 1468 rovibrational transitions) including 2046 (1106 rovibronic and 940 rovibrational transitions) from this work. The full dataset is available as electronic files as part of the \esi (these files also contain the transitions not included in the present fit).

\subsubsection{Hamiltonian}
The \pgo Hamiltonian used for the combined fit is similar to the previous studies using \pgo on \ccc. One difference with the work of \citet{Haddad2014} is that we expressed the Hamiltonian in terms of the rotational angular momentum of the nuclear framework, $\mathbf{\hat{R}}$, in order to obtain energy levels and rotational constants comparable with most of the literature data. 
It is worth noting that the specific Hamiltonian used by the \pgo software differs from the conventional Hamiltonian for a linear molecule developed for instance by \citet{Yamada1985}.
The off-diagonal constants accounting for $l$-type doubling are expressed as perturbation terms between two states, which results in several $q$ values for a given vibrational level instead of a more physically-relevant single one (see Table S4). For example, for the \xt(050) vibrational level there are three different $q$ values: one for the $\Pi$ state, and two defined as $\bra{\Pi_u}q\ket{\Phi_u}$ and $\bra{\Phi_u}q\ket{\mathrm{H}_u}$.
The resulting \pgo files together with details on their construction are provided in the \textbf{Supplementary Material}. 

The perturbation analysis for the \at(000) state is carried out using the same effective Hamiltonian as reported in \citet{Haddad2014}, with the two perturbing $\Sigma$ and $P=1$ states identified in the literature treated as three perturbing states with the $e$ and $f$ levels of the  $P=1$ state treated separately. The resulting states are labeled u$\Sigma$, u$P_e$, and u$P_f$. As in \citet{Haddad2014}, the spin-spin interaction constant $\lambda$ was fixed to 0.1 \wn in the u$\Sigma$ state. Table S5 in the \esi presents the resulting constants in these perturbing states.

\subsubsection{Fit results}

\begin{figure}[ht!]
    \centering
    \includegraphics[width=\columnwidth]{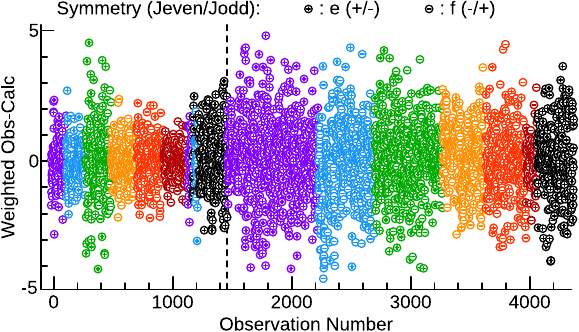}
    \caption{Residuals of the combined fit. Transitions involving the \xt(0$v_2$0) vibrational level, with up to five quanta of excitation in $\nu_2$, are plotted in a color sequence ranging from purple to dark red (same as in Figure \ref{fig:C3bands}); data in black arise from other \xt\ vibrational levels.  The vertical dashed line separates the rovibrational data (low observation numbers) from the \ax electronic data. ``Weighted Obs-Calc'' corresponds to the Obs-Calc value divided by the frequency error of the transition. }
    \label{fig:res}
\end{figure}

A total of 340 parameters have been adjusted to reproduce more than 4400 experimental data with a rms of 0.041 \wn and a reduced standard deviation of 1.3. Figure \ref{fig:res} displays the residuals of the fit with a color coding based on the \xt($v_1v_2v_3$) level involved in the transition (the same as in Figure \ref{fig:C3bands}). Overall, the fit is quite satisfactory. One can notice that the infrared (021)--(020) transitions (in green, around observation \#400) are among those the least well reproduced by the fit as a result of severe perturbations in the \xt(021) level.

The full list of parameters is reported in Tables S4 and S5 in the \esi, and a subset of parameters pertaining to the \xt(0$v_2$0) levels is reported in Table \ref{tab:CtsX0v20} where they are compared to available literature data. Only five parameters could not be determined in the present fit, the energies of five levels [\xt(110), \xt(300), \xt(400), \xt(500), \xt(600)] involved each in a single rovibrational transition for which no cross-correlating data exist.


\setlength{\tabcolsep}{3pt}

\begin{table*}[ht!]
	\scriptsize
	\centering
	\caption{Spectroscopic constants (in \wn) of \ccc in the \xt(0$v_2$0) state, with $v_2=0-5$. Values derived in this study are compared to literature data, when available. Numbers in parentheses are 1$\sigma$ deviations of the fit, in units of the last digit of the parameter (for literature data, the information is sometimes not available).}
	\label{tab:CtsX0v20}

\begin{tabular}{ccc 
S[table-format=4.11]S[table-format=1.12] S[table-format=1.7]S[table-format=1.6]S[table-format=3.6] 
S[table-format=3.9]S[table-format=3.6] S[table-format=4.5]S[table-format=3.5]S[table-format=3.5]}
\toprule
\multicolumn{2}{c}{Level} &                                  &  \multicolumn{1}{c}{$E$}                                 &  \multicolumn{1}{c}{$B$}  &  \multicolumn{1}{c}{$D\times 10^{6}$} &  \multicolumn{1}{c}{$H\times 10^{10} $}  &  \multicolumn{1}{c}{$L \times 10^{14}$} &  \multicolumn{1}{c}{$-q/2 \times 10^{3}$} &  \multicolumn{1}{c}{$-q_D/2  \times 10^{7}$} &  \multicolumn{1}{c}{$-q_H/2 \times 10^{11}$} &  \multicolumn{1}{c}{$-q_L/2 \times 10^{14}$} &  \multicolumn{1}{c}{$-q_M/2 \times 10^{18}$} \\
\midrule
$\mathrm{\Tilde{X}}$(000) &  $\Sigma_g^+$     &  This work                       &  0                                      &  0.43058717(17)   &  1.5337(25)        &  1.905(17)         &  -1.599(32)         &                       &                          &                          &                          &                          \\
                          &                   &  Ref. \cite{Kawaguchi1989}       &  0                                      &  0.430579(17)     &  1.485(22)         &  1.385(77)           &                     &                       &                          &                          &                          &                          \\
$\mathrm{\Tilde{X}}$(010) &  $\Pi_u$          &  This work                       &  63.41659112(48)                                 &  0.44241578(25)   &  2.3525(28)        &  2.675(23)         &  -2.060(53)         &  -2.850533(79)     &  5.048(19)          &  -12.14(35)         &  2.52(21)           &  -2.66(37)                \\
                          &                   &  Ref. \cite{Kawaguchi1989}$^a$   &  \multicolumn{1}{l}{~~$E_{010}$}          &  0.442381(18)     &  2.328(37)         &  2.62(11)            &                     &                       &                          &                          &                          &                          \\
$\mathrm{\Tilde{X}}$(020) &  $\Sigma_g^+$     &  This work                       &  132.80029(78)                                   &  0.4515840(62)    &  2.419(10)         &  1.571(31)           &                     &                       &                          &                          &                          &                          \\
                          &                   &  Ref. \cite{Kawaguchi1989}       &  132.7993(19)                           &  0.451632(41)     &  2.57              &                      &                     &                       &                          &                          &                          &                          \\
                          &  $\Delta_g$       &  This work                       &  133.03564(82)                                   &  0.4530995(67)    &  2.929(14)         &  3.185(88)         &  -2.52(17)          &            &                &                &                          &                          \\
                          &                   &  Ref. \cite{Kawaguchi1989}$^a$   &  133.065(29)                            &  0.453088(31)     &  2.775(50)         &  5.11(49)            &                     &    -3.788(23)       &  4.3(15)    &                          &                          &                          \\
                                &  $\braket{\Sigma_g^+}{\Delta_g}$    &  This work                                                & &                   &                    &                                                     &                    &  -3.8250(16)       &  5.074(28)          &  -5.143(89)  &                     &                     \\
$\mathrm{\Tilde{X}}$(030) &  $\Pi_g$          &  This work                       &  207.4240(11)                                    &  0.4604729(66)    &  2.6930(94)        &  1.799(45)         &  -0.465(76)         &  -5.1990(35)       &  5.928(60)          &  -5.20(20)               &                          &                          \\
                          &                   &  Ref. \cite{Gausset1965}         &  \multicolumn{1}{l}{$E_{010}+143.88$}   &  0.4600           &  2.2               &                      &                     &  -5.6                 &                          &                          &                          &                          \\
                          &  $\Phi_g$         &  This work                       &  208.3943(15)                                    &  0.4628874(98)    &  3.248(18)         &  2.79(11)          &  -1.45(23)          &            &              &                 &                          &                          \\
                                &  $\braket{\Pi_u}{\Phi_u}$        &  This work     &                                                  &                   &                    &                   &                    &  4.5111(22)        &  -5.249(43)         &  4.79(14)           &                     &                     \\
$\mathrm{\Tilde{X}}$(040) &  $\Sigma_g^+$     &  This work                       &  286.5641(13)                                    &  0.467958(12)     &  2.762(21)         &  1.585(80)           &                     &                       &                          &                          &                          &                          \\
                          &                   &  Ref. \cite{Gausset1965}         &  286.52                                 &  0.4675           &  3.248(18)         &                      &                     &                       &                          &                          &                          &                          \\
                          &  $\Delta_g$       &  This work                       &  287.2198(11)                                    &  0.4689518(86)    &  2.884(13)         &  1.569(51)           &                     &             &             &                  &  -1.17(14)               &                          \\
                          &  $\Gamma_g$       &  This work                       &  289.1579(22)                                    &  0.472027(11)     &  3.527(15)         &  2.484(52)           &                     &           &                 &                 &                          &                          \\
                                &  $\braket{\Sigma_g^+}{\Delta_g}$    &  This work &                                                  &                   &                    &                     &                    &  6.1731(30)        &  -6.804(77)         &  9.52(56)           &  -1.17(14)          &                     \\
                                &  $\braket{\Delta_g}{\Gamma_g}$   &  This work    &                                                  &                   &                    &                   &                    &  -5.0522(30)       &  5.130(65)          &  -3.88(25)          &                     &                     \\
$\mathrm{\Tilde{X}}$(050) &  $\Pi_g$          &  This work                       &  370.4479(18)                                    &  0.475305(15)     &  3.064(26)         &  3.28(10)            &                     &  -7.3000(90)       &  4.75(25)                &                          &                          &                          \\
                          &  $\Phi_g$         &  This work                       &  372.0318(17)                                    &  0.477088(16)     &  2.648(28)         &                      &                     &            &                &                          &                          &                          \\
                          &  $\mathrm{H}_g$   &  This work                       &  375.121(12)                                     &  0.480649(30)     &  4.448(64)         &  4.47(22)            &                     &            &                 &                          &                          &                          \\
                                &  $\braket{\Pi_u}{\Phi_u}$          &  This work &                                                  &                   &                    &                     &                    &  6.9550(84)        &  -8.07(24)          &                     &                     &                     \\
                                &  $\braket{\Phi_u}{\mathrm{H}_u}$   &  This work &                                                  &                   &                    &                   &                    &  -5.228(27)        &  1.56(25)           &                     &                     &                     \\
\bottomrule

\end{tabular}

\medskip
\begin{minipage}{18cm}
$^a$ Average values of the $e$ and $f$ components reported by the authors.
\end{minipage}
\end{table*}

\section{Prediction of $\nu_2$ hot bands \label{sec:prednu2}} 

The present modeling of \ccc in its electronic ground and \at\ states can be used to predict further transitions not directly observable. This is particularly interesting for the hot bands of the $\nu_2$ fundamental that remain elusive in the laboratory to date.
Table \ref{tab:predicted FIR} contains lists of predicted transitions for the \xt(020)--\xt(010) and \xt(030)--\xt(020) band systems. 
These transitions have been calculated using \pgo and the final set of parameters reported in Table \ref{tab:CtsX0v20}.
One limitation of \pgo is that the list of energies does not carry frequency error information, hence we cannot convey these errors to the transitions frequencies. However, based of the frequency errors of the transitions used to determine these energies, and the overall good quality of the fit, we estimate the accuracy of these transition frequencies to be of the order of 0.005~\wn (150 MHz).

\begin{table*}\label{tab:predicted FIR}
\scriptsize
\centering
\caption{Low-$J$ far-infrared transitions of the \xt(020)--\xt(010) and \xt(030)--\xt(020) hot bands of $\nu_2$ predicted using our model (in~cm$^{-1}$). The $e$/$f$ label of the lower state is reported.} 
\begin{tabular}{l ccc c ccc c ccc}
\toprule
\textbf{$J''$} & \textbf{P} & \textbf{Q} & \textbf{R} && \textbf{P} & \textbf{Q} & \textbf{R} && \textbf{P} & \textbf{Q} & \textbf{R}\\
\midrule

&\multicolumn3c{(020)--(010) $\Sigma-\Pi$} && \multicolumn3c{(020)--(010) $\Delta-\Pi$}\\ \cmidrule{2-4} \cmidrule{6-8}
1 & 68.947($e$) &             & 71.657($e$) &&                   &                   & \rev{70.088($e$)}&&&\\
2 &             & 69.864($f$) &             &&                   & 68.296\rev{($f$)} & \rev{71.015($f$)}&&&\\
3 & 67.261($e$) &             & 73.589($e$) && 65.693\rev{($e$)} & 68.412\rev{($e$)} & \rev{72.029($e$)}&&&\\
4 &             & 69.959($f$) &             && 64.782\rev{($f$)} & 68.399\rev{($f$)} & \rev{72.936($f$)}&&&\\
5 & 65.678($e$) &             & 75.634($e$) && 64.118\rev{($e$)} & 68.655\rev{($e$)} & \rev{74.058($e$)}&&&\\
6 &             & 70.123($f$) &             && 63.144\rev{($f$)} & 68.547\rev{($f$)} & \rev{74.918($f$)}&&&\\

\midrule
& \multicolumn3c{(030)--(020) $\Pi-\Sigma$}  && \multicolumn3c{(030)--(020) $\Pi-\Delta$}  && \multicolumn3c{(030)--(020) $\Phi-\Delta$} \\ \cmidrule{2-4} \cmidrule{6-8} \cmidrule{10-12}
0 &        &        & 75.074($e$) &&                 &                 &                 &&& \\
1 &        &        &        &&                 &                 &                 &&& \\
2 & 72.364($e$) & 74.247($e$) & 76.917($e$) && 73.933\rev{($e$)} & 75.816\rev{($e$)} & 78.486\rev{($e$)} &&                 &           & 75.841\rev{($e$)} \\
3 &        &        &        && 73.097\rev{($f$)} & 75.767\rev{($f$)} & 79.617\rev{($f$)} &&                 & 73.122\rev{($f$)} & 76.824\rev{($f$)} \\
4 & 70.589($e$) & 74.440($e$) & 78.788($e$) && 72.149\rev{($e$)} & 76.000\rev{($e$)} & 80.348\rev{($e$)} && 69.505\rev{($e$)} & 73.206\rev{($e$)} & 77.830\rev{($e$)} \\
5 &        &        &        && 71.463\rev{($f$)} & 75.811\rev{($f$)} & 81.712\rev{($f$)} && 68.670\rev{($f$)} & 73.294\rev{($f$)} & 78.842\rev{($f$)} \\
6 & 68.831($e$) & 74.732($e$) & 80.682($e$) && 70.408\rev{($e$)} & 76.309\rev{($e$)} & 82.258\rev{($e$)} && 67.890\rev{($e$)} & 73.438\rev{($e$)} & 79.902\rev{($e$)} \\

\bottomrule
\end{tabular}
\end{table*}

\section{Astronomical implications \label{sec:astro}} 

Due to the extremely low bending frequency of \ccc , even in an  environment at moderate temperature its low excited bending vibrational levels \xt(010), at $\sim$ 63 \wn (91\,K),  and \xt(020), at $\sim 132$ \wn (191\,K), can be thermally populated significantly.
Indeed, at 100\,K, 29\,\% and 13\,\% of the  GS population lies in each of these levels; at 50\,K, these numbers drop down to 14\,\% and 2\,\%, which remains significant for an abundant molecule. Hence, $\nu_2$ hot bands may be detectable in various environments of the interstellar medium where \ccc is abundant.
Unambiguous detections, $i.e.$, beyond the line confusion limit, become possible when astronomical data can be compared to accurate submillimeter laboratory data (with sub-MHz resolution) that will strongly benefit from the predictions made here. Given the accurate parameters derived here, even without such laboratory data it should be possible to identify these $\nu_2$ excited transitions in astronomical data.
The 0.005 \wn (150 MHz) accuracy of our predictions corresponds to a velocity uncertainty of $ \Delta V $ = 23.6\,km$\cdot$s$^{-1} $. 
Since the linewidth of observed spectra of \ccc $\nu_2$ fundamental band in star forming region ranges from 5 to 12\,km$\cdot$s$^{-1} $  \citep{Giesen2001, Mookerjea2010, Mookerjea2012,Giesen2020}, we conclude that our data can be used to search for \ccc $\nu_2$ hot bands. 
Measurement of \ccc in excited states and determination of its abundance and excitation temperature may give new insights into the chemistry of its formation and will add further information to derive the origin of small and possibly also longer carbon chains in the ISM.

\section{Conclusion} \label{sec:conc} 
This work presents the most complete spectroscopic study of \ccc to date.
Through the combination of infrared and optical transitions precise vibrational energies and rotational constants for the low-lying bending modes of \ccc are determined up to $v_2 = 5$, significantly extending our knowledge of the rovibrational manifold of the electronic GS. The measured and predicted transition frequencies presented here can be used in astronomical observations in both the optical, infrared, and far-infrared, increasing the effectiveness of \ccc as a probe of the physical and chemical environment of the target.  \pgo files allowing the full simulation of all known bands of \ax system are given in the \esi so that further improvements of the laboratory spectroscopy of this molecule can be directly incorporated in the model.

\section{Acknowledgments}
This manuscript comprises of two datasets. The project started with LIF measurements at USTC (recorded in 2017) and upon analysis it became clear that combining these with non-published infrared measurements recorded at SOLEIL in 2010 offered a unique opportunity for a very complete spectroscopic study of the \ccc radical. PGOPHER is ideal to merge the two large data sets. Given all involved spectroscopic challenges, we asked Colin Western for help, and as usual were helped on the spot and far beyond, resulting in his co-authorship. Along the way of finishing this manuscript, Colin sadly passed away. We dedicate this manuscript to his memory.\\
We acknowledge financial support 
from the National Natural Science Foundation of China (22173089 and 21827804), the Netherlands Organization for Scientific Research (NWO) through a VICI grant, the Netherlands Research School for Astronomy (NOVA), and the \textit{Programme National ``Physique et Chimie du Milieu Interstellaire''} (PCMI) of CNRS/INSU with INC/INP co-funded by CEA and CNES. DT acknowledges funding from the Natural Sciences and Engineering Research Council of Canada in the form of a Discovery grant. 
Part of this work was performed at the SOLEIL facility under the proposal 20100296. 

\section{Authors contributions}

\textbf{Marie-Aline Martin-Drumel:} Investigation, Formal analysis, Writing -- original draft; Writing -- review \& editing;
\textbf{Qiang Zhang:} Investigation, Formal analysis, Methodology, Writing – original draft, Writing –- review \& editing;
\textbf{Kirstin Doney:} Investigation, Formal analysis, Writing -- original draft; Writing -- review \& editing;
\textbf{Olivier Pirali:} Conceptualization, Investigation, Formal analysis, Writing -- original draft; Writing -- review \& editing;
\textbf{Michel Vervloet:} Investigation, Formal analysis, Writing -- review \& editing;
\textbf{Dennis Tokaryk:} Conceptualization, Investigation, Writing -- review \& editing;
\textbf{Colin Western:} Software;
\textbf{Harold Linnartz: } Resources, Supervision, Writing -- review \& editing;
\textbf{Yang Chen:} Conceptualization, Investigation; Supervision; Writing -– review \& editing
\textbf{Dongfeng Zhao:} Conceptualization, Investigation, Supervision, Writing –- review \& editing.

\bibliographystyle{elsarticle-num-names} 
\biboptions{sort&compress}
\bibliography{biblio}

\end{document}